\documentclass[aps,prl,reprint,superscriptaddress,showpacs,showkeys]{revtex4-1}

\usepackage[dvipdfmx]{graphicx}
\usepackage{lineno}
\usepackage{color}
\usepackage{comment}
\usepackage{amsmath}

\begin{document}

%\linenumbers

%\preprint{}

%\title{First direct mass measurements of hot-fusion transuranium isotopes with an MRTOF-MS}
%\title{First direct mass measurements of \textcolor{red}{mendelevium} with an MRTOF-MS}
%\title{First direct mass measurements of nuclides around $Z=100$ with an MRTOF-MS}
%\title{First Direct Mass Measurements of Nuclides around $Z=100$ with a Multi-Reflection Time-of-Flight Mass Spectrograph}
\title{First Direct Mass Measurements of Nuclides around $Z=100$ with a Multireflection Time-of-Flight Mass Spectrograph}


\author{Y.~Ito}
% \email[]{yito@riken.jp}
\email[Present E-mail address: ]{yito@physics.mcgill.ca}
%\thanks{}
\altaffiliation[Present address: ]{Department of Physics, MacGill University, Montreal, Quebec H3A 2T8, Canada}
\affiliation{RIKEN Nishina Center for Accelerator-Based Science, Wako 351-0198, Japan}
\author{P.~Schury}
\affiliation{Wako Nuclear Science Center (WNSC), Institute of Particle and Nuclear Studies (IPNS),
High Energy Accelerator Research Organization (KEK), Wako 351-0198, Japan}
\author{M.~Wada}
\affiliation{RIKEN Nishina Center for Accelerator-Based Science, Wako 351-0198, Japan}
\affiliation{Wako Nuclear Science Center (WNSC), Institute of Particle and Nuclear Studies (IPNS),
High Energy Accelerator Research Organization (KEK), Wako 351-0198, Japan}
\author{F.~Arai}
\affiliation{RIKEN Nishina Center for Accelerator-Based Science, Wako 351-0198, Japan}
\affiliation{University of Tsukuba, Tsukuba, Ibaraki 305-8577, Japan}
\author{H.~Haba}
\affiliation{RIKEN Nishina Center for Accelerator-Based Science, Wako 351-0198, Japan}
\author{Y.~Hirayama}
\affiliation{Wako Nuclear Science Center (WNSC), Institute of Particle and Nuclear Studies (IPNS),
High Energy Accelerator Research Organization (KEK), Wako 351-0198, Japan}
\author{S.~Ishizawa}
\affiliation{RIKEN Nishina Center for Accelerator-Based Science, Wako 351-0198, Japan}
\affiliation{Graduate School of Science and Engineering, Yamagata University, Yamagata 990-8560, Japan}
\author{D.~Kaji}
\affiliation{RIKEN Nishina Center for Accelerator-Based Science, Wako 351-0198, Japan}
\author{S.~Kimura}
\affiliation{RIKEN Nishina Center for Accelerator-Based Science, Wako 351-0198, Japan}
\affiliation{Wako Nuclear Science Center (WNSC), Institute of Particle and Nuclear Studies (IPNS),
High Energy Accelerator Research Organization (KEK), Wako 351-0198, Japan}
\affiliation{University of Tsukuba, Tsukuba, Ibaraki 305-8577, Japan}
\author{H.~Koura}
\affiliation{Japan Atomic Energy Agency, Tokai, Ibaraki 319-1185, Japan}
\author{M.~MacCormick}
\affiliation{Institut de Physique Nucl\'eaire, IN2P3-CNRS, Universit\'e Paris-Sud,
Universit\'e Paris-Saclay, 91406 Orsay Cedex, France}
%Institut de Physique Nucl\UTF{00E9}aire, IN2P3-CNRS, Universit\UTF{00E9} Paris-Sud, Universit\UTF{00E9} Paris-Saclay, 91406 Orsay Cedex, France
\author{H.~Miyatake}
\affiliation{Wako Nuclear Science Center (WNSC), Institute of Particle and Nuclear Studies (IPNS),
High Energy Accelerator Research Organization (KEK), Wako 351-0198, Japan}
\author{J.Y.~Moon}
\affiliation{Wako Nuclear Science Center (WNSC), Institute of Particle and Nuclear Studies (IPNS),
High Energy Accelerator Research Organization (KEK), Wako 351-0198, Japan}
\affiliation{Rare Isotope Science Project, Institute for Basic Science (IBS), Daejeon 305-811, Korea}
\author{K.~Morimoto}
\affiliation{RIKEN Nishina Center for Accelerator-Based Science, Wako 351-0198, Japan}
\author{K.~Morita}
\affiliation{RIKEN Nishina Center for Accelerator-Based Science, Wako 351-0198, Japan}
\affiliation{Department of Physics, Kyushu University, Nishi-ku, Fukuoka 819-0395, Japan}
\author{M.~Mukai}
\affiliation{Wako Nuclear Science Center (WNSC), Institute of Particle and Nuclear Studies (IPNS),
High Energy Accelerator Research Organization (KEK), Wako 351-0198, Japan}
\affiliation{University of Tsukuba, Tsukuba, Ibaraki 305-8577, Japan}
\author{I.~Murray}
\affiliation{RIKEN Nishina Center for Accelerator-Based Science, Wako 351-0198, Japan}
\author{T.~Niwase}
\affiliation{Department of Physics, Kyushu University, Nishi-ku, Fukuoka 819-0395, Japan}
\author{K.~Okada}
\affiliation{RIKEN Nishina Center for Accelerator-Based Science, Wako 351-0198, Japan}
\affiliation{Sophia University, Chiyoda-ku, Tokyo 102-8554, Japan}
\author{A.~Ozawa}
\affiliation{University of Tsukuba, Tsukuba, Ibaraki 305-8577, Japan}
\author{M.~Rosenbusch}
\affiliation{RIKEN Nishina Center for Accelerator-Based Science, Wako 351-0198, Japan}
\author{A.~Takamine}
\affiliation{RIKEN Nishina Center for Accelerator-Based Science, Wako 351-0198, Japan}
\author{T.~Tanaka}
\affiliation{RIKEN Nishina Center for Accelerator-Based Science, Wako 351-0198, Japan}
\affiliation{Department of Physics, Kyushu University, Nishi-ku, Fukuoka 819-0395, Japan}
\author{Y.X.~Watanabe}
\affiliation{Wako Nuclear Science Center (WNSC), Institute of Particle and Nuclear Studies (IPNS),
High Energy Accelerator Research Organization (KEK), Wako 351-0198, Japan}
\author{H.~Wollnik}
\affiliation{RIKEN Nishina Center for Accelerator-Based Science, Wako 351-0198, Japan}
\affiliation{Department of Chemistry and Biochemistry, New Mexico State University, Las Cruces, New Mexico 88003,USA}
\author{S.~Yamaki}
\affiliation{RIKEN Nishina Center for Accelerator-Based Science, Wako 351-0198, Japan}
\affiliation{Department of Physics, Saitama University, Sakura-ku, Saitama 338-8570, Japan}


\date{\today}

\begin{abstract}

The masses of $^{246}$Es, $^{251}$Fm and the transfermium nuclei $^{249-252}$Md, and $^{254}$No, produced by hot- and cold-fusion reactions, in the vicinity of the deformed $N=152$ neutron shell closure, have been directly measured using a multireflection time-of-flight mass spectrograph.  The masses of $^{246}$Es and $^{249,250,252}$Md were measured for the first time.  Using the masses of $^{249,250}$Md as anchor points for $\alpha$ decay chains, the masses of heavier nuclei, up to $^{261}$Bh and $^{266}$Mt, were determined.  These new masses were compared with theoretical global mass models and demonstrated to be in good agreement with macroscopic-microscopic models in this region.  The empirical shell gap parameter $\delta_{2n}$ derived from three isotopic masses was updated with the new masses and corroborate the existence of the deformed $N=152$ neutron shell closure for Md and Lr.
 
\end{abstract}

%\keywords{}

\maketitle


%\section{Introduction}

\par %\textcircled{\scriptsize 1}
Precision mass measurements of unstable nuclei, providing a direct measure of the nuclear binding energy, are invaluable for the study of nuclear shell evolution and collective effects, such as deformations, far from stability \cite{Lunney:2003ue, SORLIN2008a}. For transfermium nuclei and the yet poorly investigated region towards the superheavy nuclei (SHN), where proton repulsion becomes a generally dominant feature, the description of nuclear lifetimes depends crucially on shell stabilization effects mainly driven by deformed shells \cite{SOBICZEWSKI1966500, PATYK1989267, PATYK1991132}. Theoretical studies with increasing particle numbers investigate the so-called ``island of stability'' \cite{seaborg1987}, where features like the continuing decrease of energy gaps \cite{BENDER200142} and the emergence of shape coexistence \cite{Cwiok2005} have crucial impact on the predicted position and localization of stability regions and the corresponding lifetimes of the nuclei. Although first experimental evidence for SHN has reached the region of the predicted sub-shell closure at $N=162$ \cite{PhysRevC.54.620, Oganessian:2011cm, Rudolph2013}, the deformed shell closure at $N=152$ for transfermium nuclei (see \emph{e.g.}~\cite{JULIN2001221}) and, as recently pointed out, weaker shell effects in the vicinity \cite{PhysRevC.87.054324}, still represents the cutting edge for thorough experimental investigations.  The transfermium nuclei, however, can only be produced online, in heavy-ion fusion and nucleon transfer reactions, and consequently only low yields are available for study, necessitating highly efficient techniques.  Direct mass measurements of transfermium nuclei have so far been performed for only six nuclei -- four isotopes of nobelium and two isotopes of lawrencium -- with the Penning trap mass spectrometer SHIPTRAP \cite{block2010785, ramirez20121207}.

\begin{figure}
\includegraphics[width=\linewidth]{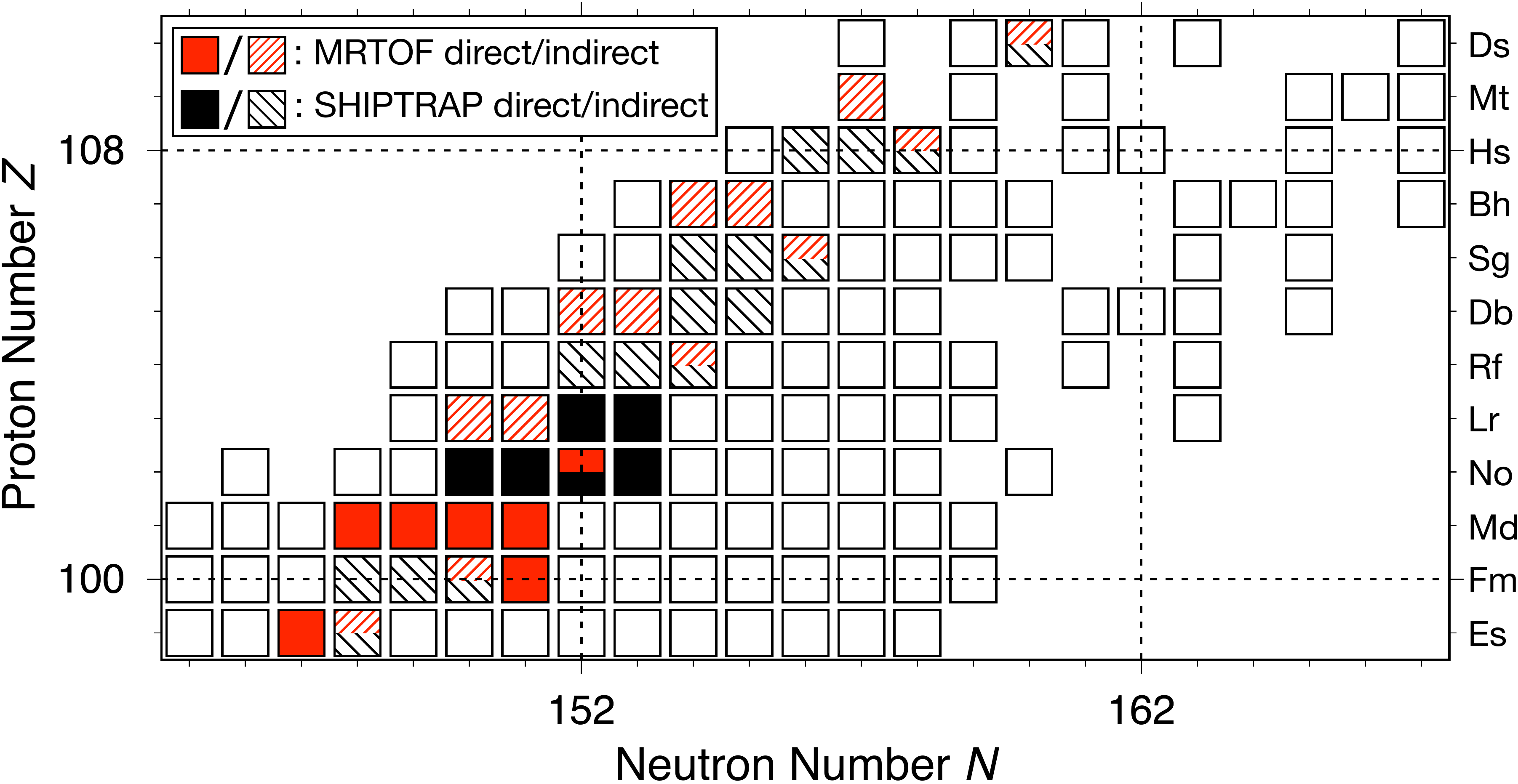}%
\caption{Nuclear chart above Californium ($Z=98$). The squares indicate nuclei synthesized so far. Nuclei whose masses were determined in this work are indicated by the solid (direct) and the left-hatched (indirect) red squares. Similarly, nuclei whose masses were determined by SHIPTRAP measurements are indicated by the solid (direct) and the right-hatched (indirect) black squares.}
\label{fig:chart}
\end{figure}

\par %\textcircled{\scriptsize 2}
In this letter we report the first implementation of a multireflection time-of-flight mass spectrograph (MRTOF-MS) for transfermium nuclei as shown in Fig.~\ref{fig:chart}, including new mass measurements of $^{246}$Es, $^{251}$Fm, $^{249-252}$Md, and $^{254}$No, performed with sub-ppm precision.  This represents the first determination of the masses of $^{249-250,252}$Md, closing a gap of unmeasured nuclei which could not be linked by corresponding decay chains.  Through combining the masses of the dominantly $\beta$-decaying nuclei $^{249-250}$Md with previously-known $Q_\alpha$-values, the masses of nuclei up to $^{261}$Bh and $^{266}$Mt could be experimentally determined for the first time.  These results support the existence of the $N=152$ shell gap in Md and Lr, while also providing the first experimental data at the shell gap for Db isotopes.

%\section{Experiment}

\begin{figure}
\includegraphics[width=\linewidth]{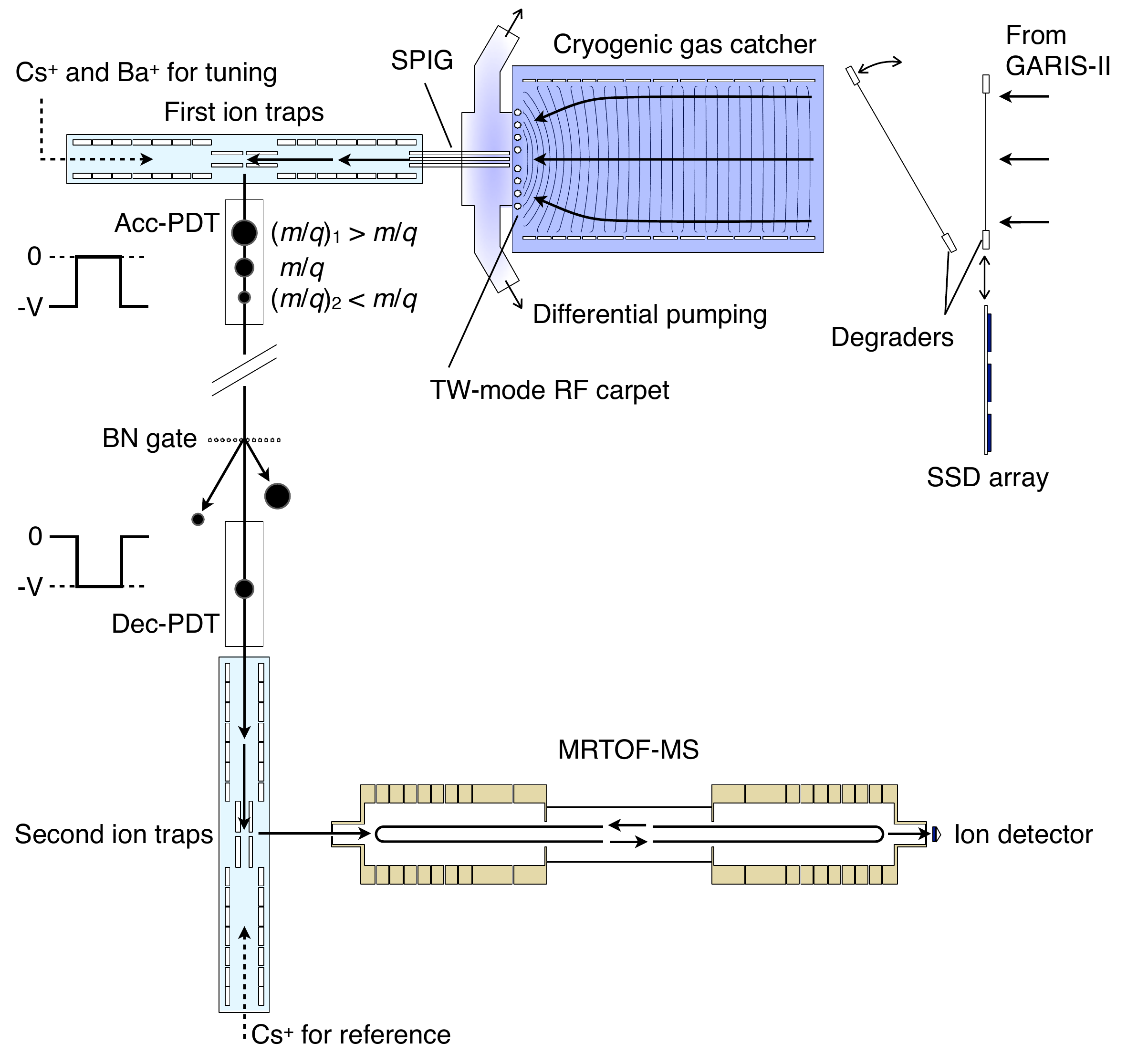}%
\caption{Schematic view of the experimental setup. Dashed and solid arrows schematically depict the paths of reference and analyte ions, respectively. Curved lines in the cryogenic gas cell represent equipotential lines of the static field that focus ions onto the rf carpet, while arrows depict the nominal trajectories of ions in the static field.}
\label{fig:gc-mtof}
\end{figure}

\par %\textcircled{\scriptsize 3}
As shown in Fig.~\ref{fig:gc-mtof}, the MRTOF-MS \cite{ito2013011306,schury2014nim} was installed behind a cryogenic helium gas cell and ion trap system coupled with the gas-filled recoil ion separator GARIS-II \cite{kaji2013311}.
Primary beams provided by the RIKEN heavy-ion linear accelerator RILAC impinged upon a rotating target to produce fusion-evaporation residues (ER).
The stopping of high-energy ER in the gas cell was optimized by adjusting the thickness of a Mylar degrader while the gas cell was filled with 150~mbar helium at a temperature of 150~K.  The ions were transported to a radiofrequency carpet (RFC) \cite{wada2003570}, located on the exit wall, by a static electric field and then extracted by means of a traveling-wave (TW) mode RFC technique \cite{bollen2011131,arai201456,arai2015030110}.  The extracted ions were transported through a differential pumping section by a sextupole ion guide (SPIG) and then accumulated in the first ion trap system.
%The ion trap systems, consisting of a pair of linear Paul traps on either side of a ``flat'' ion trap \cite{ito2013544}, were filled with helium buffer gas at $\sim$10$^{-2}$~mbar.  The fore and aft linear Paul traps accumulated the continuous ion beams from the gas cell and an offline \textcolor{red}{thermal} ion source \textcolor{red}{for beam-line tuning}, respectively, in order to pre-cool and pre-bunch them prior to their transfer to the flat trap.  After accumulating and cooling in the flat trap, ion bunches were orthogonally ejected, accelerated to a kinetic energy of $\approx$1.7~keV by a pulsed drift tube (Acc-PDT), transported through an electrostatic multipole lens and a Bradbury-Nielsen gate (BN gate) \cite{bradbury1936388} and decelerated to tens of electron-volts by a second pulsed drift tube (Dec-PDT) before being retrapped in the second ion trap system, located in the experimental room underneath GARIS-II.  The two ion trap systems have identical geometries.
After accumulating and cooling in the flat trap, ion bunches were orthogonally ejected, accelerated to a kinetic energy of $\approx$1.7~keV by a pulsed drift tube (Acc-PDT), transported through an electrostatic multiple lens and a Bradbury-Nielsen gate (BN gate) \cite{bradbury1936388} and decelerated to tens of electron-volts by a second pulsed drift tube (Dec-PDT) before being retrapped in the second ion trap system, located in the experimental room underneath GARIS-II.

\par %\textcircled{\scriptsize 4}
The first and second ion trap systems, each consisting of a pair of linear Paul traps on either side of a ``flat" ion trap \cite{ito2013544}, have the same geometry and were filled with helium buffer gas at $\sim$10$^{-2}$~mbar.
In the first trap system, the fore and aft linear Paul traps accumulated the continuous ion beams from the gas cell and from a thermal ionization ion source, respectively, in order to pre-cool and pre-bunch the ion beam prior to transfer to the flat trap.
The thermal ionization ion source in the first trap system provided both Cs$^{+}$ and Ba$^{+}$ ions for beam-line tuning.
In the second trap system, the fore and aft linear Paul traps accumulated and pre-cooled the pulsed beam delivered from the first trap and the continuous beam from a reference ion source, respectively.
The second trap system's thermal ionization ion source provided reference Cs$^{+}$ ions for the mass measurements.

\begin{figure}
\includegraphics[width=\linewidth]{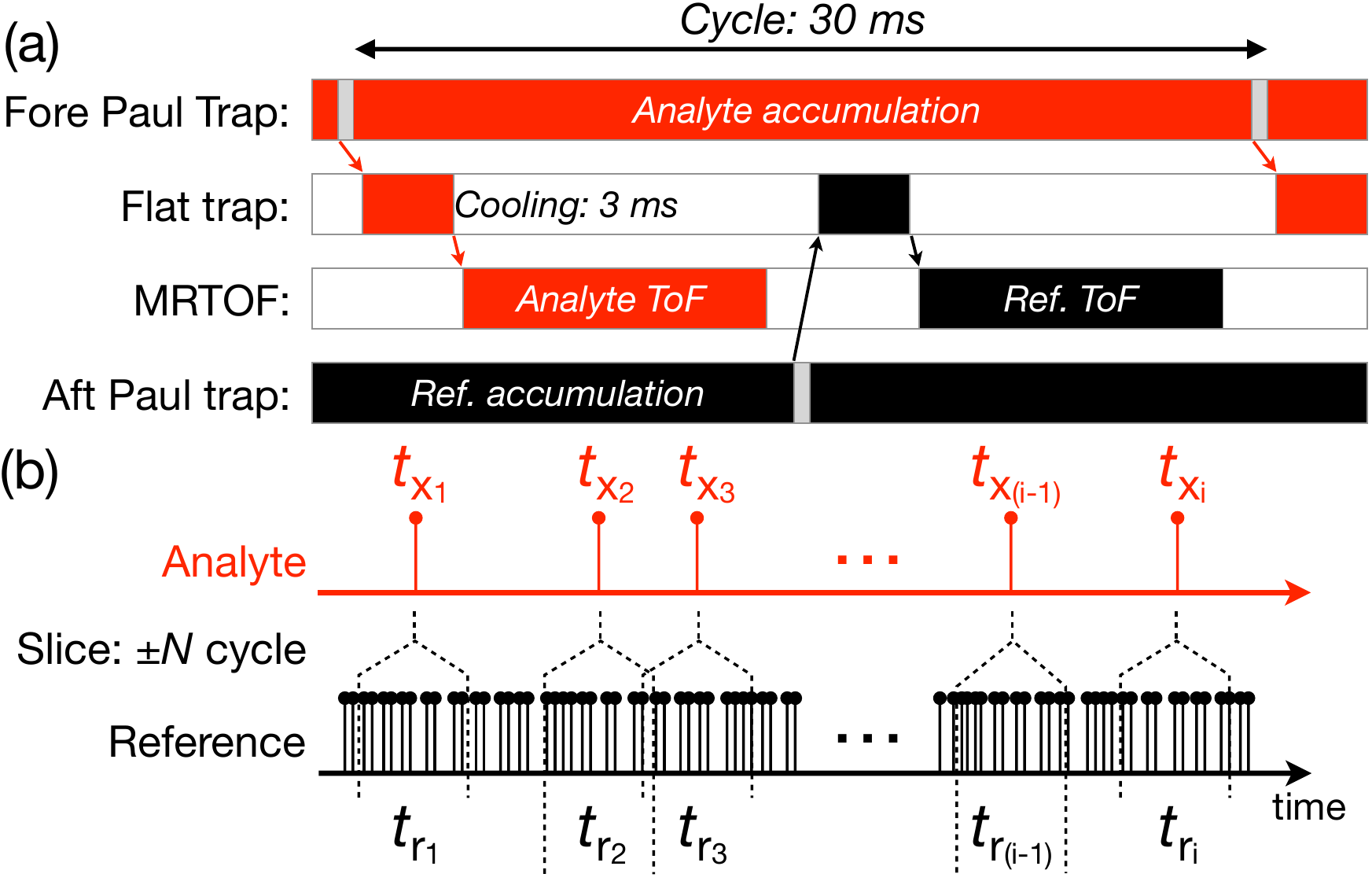}
\caption{(a) Pictorial representations of concomitant referencing scheme in the second ion trap system. (b) ``Slice-by-Event'' analysis method. }
\label{fig:concomitant}
\end{figure}

\par %\textcircled{\scriptsize 5}
The novel flat trap geometry allowed implementation of a concomitant measurement scheme, shown in Fig.~\ref{fig:concomitant}(a).  While ions from the gas cell were being analyzed with the MRTOF-MS, reference ions stored in the second trap system's aft linear Paul trap were transferred to the flat ion trap and cooled, while pulses of ions sent from the first trap system continue to accumulate in the lower trap system's fore linear Paul trap.
%while ions from the gas cell were continuously accumulated and stored in the top linear Paul trap.
In this way, measurements of analyte ions from the gas cell were interleaved with measurements of reference ions within a 30~ms cycle (15~ms for each).  In addition to providing a nearly 100\% duty cycle, the times-of-flight (ToF) of the reference ions provide precise corrections of ToF drifts for both reference ions and analyte ions.  

%\section{Data analysis}

\par %\textcircled{\scriptsize 6}
The ToF drift correction was performed in a manner we refer to as the ``Slice-by-Event'' method (see Fig.~\ref{fig:concomitant}(b)).  Because analyte detection events were sparse, it was not necessary to consider all reference events.  Rather, the reference ions detected 50 cycles (1.5 s) before and after each analyte detection event were combined to produce a reference spectrum.  The centroid of reference events in each slice was used to determine the reference ToF $t_{\mathrm{r}_{i}}$ for each analyte ToF $t_{\mathrm{x}_{i}}$.  Drift-corrected spectra can then be produced for reference and analyte by multiplying the ToF of each detected ion in subset $i$ by $t_\mathrm{r_{0}}/t_{\mathrm{r}_{i}}$.  A detailed review of this analytical method will be provided in a future publication.  Spectra were fitted with an unbinned maximum-likelihood estimator using an asymmetric combined Gaussian-Lorentzian function \cite{stancik2008}.

\par
\begin{figure}
\includegraphics[width=\linewidth]{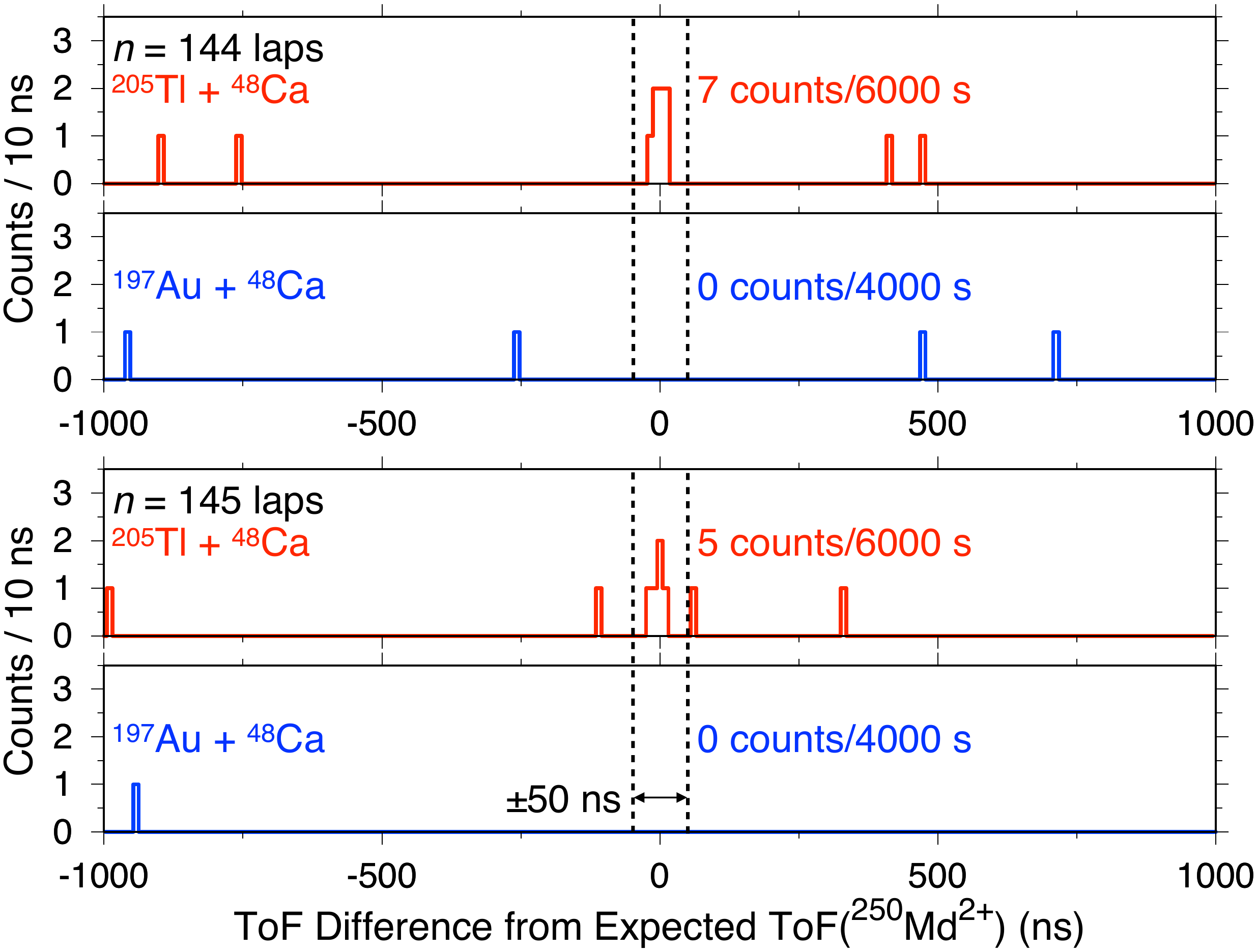}
\caption{ToF spectra in the anticipated vicinity of $^{250}$Md$^{2+}$ at $n= 144$ and 145~laps for $^{\rm nat}$Tl and $^{197}$Au targets.}
\label{fig:multi-hist}
\end{figure}

\begin{figure}
\includegraphics[width=\linewidth]{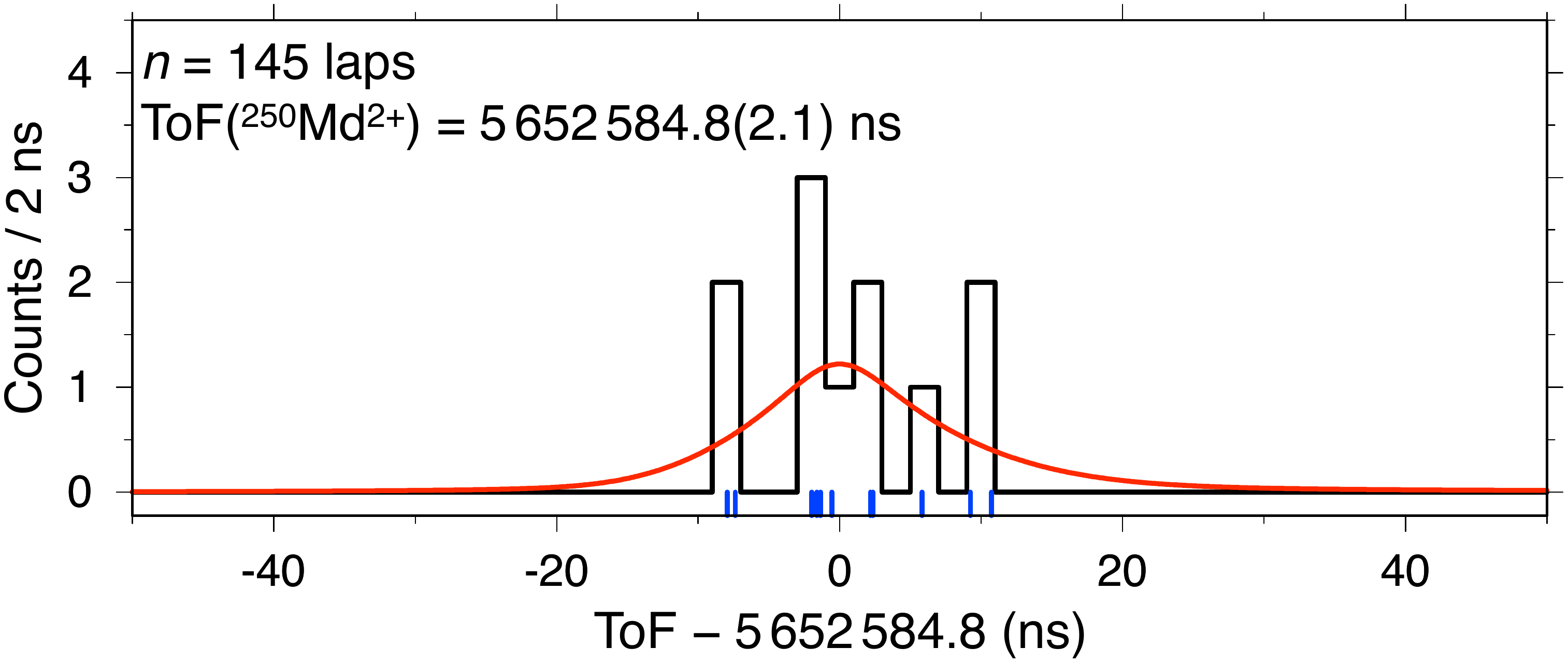}
%\caption{ToF spectrum of $^{250}$Md$^{2+}$ with fitting curve for a region of $\pm$50~ns. The shape parameters of the fitting function were pre-determined with a high statistics reference ($^{133}$Cs$^+$) peak in the same measurement.}
\caption{Fitted ToF spectrum of $^{250}$Md$^{2+}$ at $n=145$~laps. The shape parameters of the fitting function were pre-determined with a high statistics reference ($^{133}$Cs$^+$) peak in the same measurement, at the same number of laps.}
\label{fig:ToF_fitting}
\end{figure}

 %\textcircled{\scriptsize 7}
Due to the multireflection nature of the MRTOF-MS, there is not a one-to-one correspondence between ToF and $A/q$; unambiguous identification cannot be made from a single spectrum.  This is a consequence of the possibility that two ion species differing in mass-to-charge ratio by $\Delta A/q$ will also differ in number of laps made in the MRTOF-MS by $\Delta n$~laps such that they have essentially the same ToF.  To avoid misidentifications, therefore, we employed confirmation measurements of each analyte ion at different numbers of laps (generally $\pm1$ laps), and additionally for low count-rate measurements of $^{249,250}$Md further confirmations with a dummy target of lower $Z$ which is unable to produce the desired ER but that can be presumed to provide otherwise similar conditions.  Figure~\ref{fig:multi-hist} demonstrates this process in the case of $^{250}$Md.  After 6000~s, within $\pm$50~ns of the expected ToF of $^{250}$Md$^{2+}$ 7 and 5 counts, respectively, were observed at $n= 144$ and $n= 145$~laps while using $^{\rm{nat}}$Tl ($Z=81$) targets; no counts were observed when using $^{197}$Au ($Z=79$) targets in 4000~s for both.
This provides strong evidence that the observed spectral peak truly belongs to $^{250}$Md$^{2+}$ with the probability of no detected events being only 0.03\%.  
The raw and binned spectrum observed for $^{250}$Md at $n=145$ laps, along with the resultant fitting curve by an unbinned maximum-likelihood routine \cite{r2008}, is shown in Fig.~\ref{fig:ToF_fitting}. 
This process was employed for each isotope measured.
%The systematic uncertainties were evaluated from possible uncertainty in the origin of time-of-flight \cite{ito2013011306}.

\par %\textcircled{\scriptsize 8}
Experimental conditions, reactions, and primary beam energies, for each measured isotopes are included in Table~\ref{tab:mass_excess}.
%\textcolor{red}{Primary beams were provided by the RIKEN heavy-ion linear accelerator RILAC.}
$^{246}$Es, $^{251}$Fm and $^{252}$Md were produced with hot-fusion reactions using $^{18}$O and $^{19}$F primary beams with intensities of $\sim$3~p$\mu$A on $^{232}$Th and $^{\mathrm{nat}}$U targets.  A $^{48}$Ca primary beam of $\sim$0.3~p$\mu$A intensity was used to produce $^{249,250,251}$Md and $^{254}$No via cold-fusion reactions with $^{\mathrm{nat}}$Tl and $^{208}$Pb targets.  All targets had a thickness of $\sim$500~$\mu$g/cm$^{2}$ with 1.4-mg/cm$^{2}$ Ti backings for actinide targets and 60-$\mu$g/cm$^{2}$ C backing for other targets, and were mounted on a 300-mm wheel \cite{kaji201511} which rotated at 2000~rpm during irradiation.
%Cross sections for $^{246}$Es and $^{249,250,252}$Md were estimated using a Monte Carlo calculation \cite{NRVmisc}.

\begin{table*}[]
\caption{Measured isotopes, reactions, reaction energies at target center in laboratory frame ($E_\mathrm{lab}$), recoil energies (E$_\mathrm{recoil}$), cross sections ($\sigma_\mathrm{ER}$), squares of analyte-reference ToF ratio ($\rho^{2}$), mass excesses from this work ($ME_\mathrm{MRTOF}$) and from the atomic mass evaluation 2016 (AME16) ($ME_\mathrm{AME16}$) \cite{huand2017030002}, mass deviations ($\Delta m = ME_\mathrm{MRTOF}-ME_\mathrm{AME16}$), and the total number of detected ions ($N_{\rm ion}$) in this work. Parenthetical values of $\sigma_{\mathrm{ER}}$ denote estimated values from a Monte Carlo code \cite{NRVmisc}.
Extrapolated values of $ME_\mathrm{AME16}$ denoted by \#. Experimental statistical and systematic uncertainties are described in the first and second parentheses in $ME_\mathrm{MRTOF}$ and $\Delta m$, respectively.}
\label{tab:mass_excess}
\begin{ruledtabular}
\begin{tabular}{cccccccccccccc}
Isotope & Reaction  & $E_\mathrm{lab}$ & $E_\mathrm{recoil}$ & $\sigma_\mathrm{ER}$ & $\rho^{2}$     & $ME_{\rm MRTOF}$ & $ME_{\rm AME16}$       & $\Delta m$    & $N_{\rm ion}$ \\
             &                 & (MeV)                    & (MeV)                       & (nb)                                &                       & (keV/$c^{2}$)            & (keV/$c^{2}$)                            & (keV/$c^{2}$) & (counts)        \\ \hline
$^{246}$Es  & $^{232}$Th($^{19}$F, 5$n$)  & 99.6, 103 & 7.5, 7.8 & (800) \cite{NRVmisc}             & 0.92574351(44) & 67812(109)(32) & 67900$^{\#}$(224$^{\#}$) & -88(109)(32)  & 33 \\
$^{251}$Fm  & $^{238}$U($^{18}$O, 5$n$)   & 93.9      & 6.9 & 4000 \cite{donets1966}         & 0.94458700(14) & 75996(34)(25)  & 75954(15)                & 42(34)(25)    & 397 \\
$^{249}$Md  & $^{203}$Tl($^{48}$Ca, 2$n$) & 215       & 41.1 & (40) \cite{NRVmisc}        & 0.93706792(89) & 77259(221)(26) & 77232$^{\#}$(205$^{\#}$) & 27(221)(26)   & 14 \\
$^{250}$Md  & $^{205}$Tl($^{48}$Ca, 3$n$) & 223       & 42.3 & (200) \cite{NRVmisc}       & 0.94083491(56) & 78472(138)(25) & 78630$^{\#}$(298$^{\#}$) & -158(138)(25) & 29 \\
$^{251}$Md  & $^{205}$Tl($^{48}$Ca, 2$n$) & 215       & 40.8 & 760 \cite{chatillon2007132503} & 0.94459923(24) & 79025(60)(23)  & 78967(19)                & 58(60)(23)    & 173  \\
$^{252}$Md  & $^{238}$U($^{19}$F, 5$n$)   & 98.6      & 7.3 & (500) \cite{NRVmisc}             & 0.94836715(36) & 80467(89)(22)  & 80511$^{\#}$(130$^{\#}$) & -44(89)(22)   & 63 \\
$^{254g}$No & $^{208}$Pb($^{48}$Ca, 2$n$) & 219       & 41.1 & 2000 \cite{oganessian2001}     & 0.95590832(17) & 84675(42)(19)  & 84723.4(9.3)             & -48(42)(19)   & 398    
\end{tabular}
\end{ruledtabular}
\end{table*}

%\section{Results}

\par
\begin{figure}
\includegraphics[width=\linewidth]{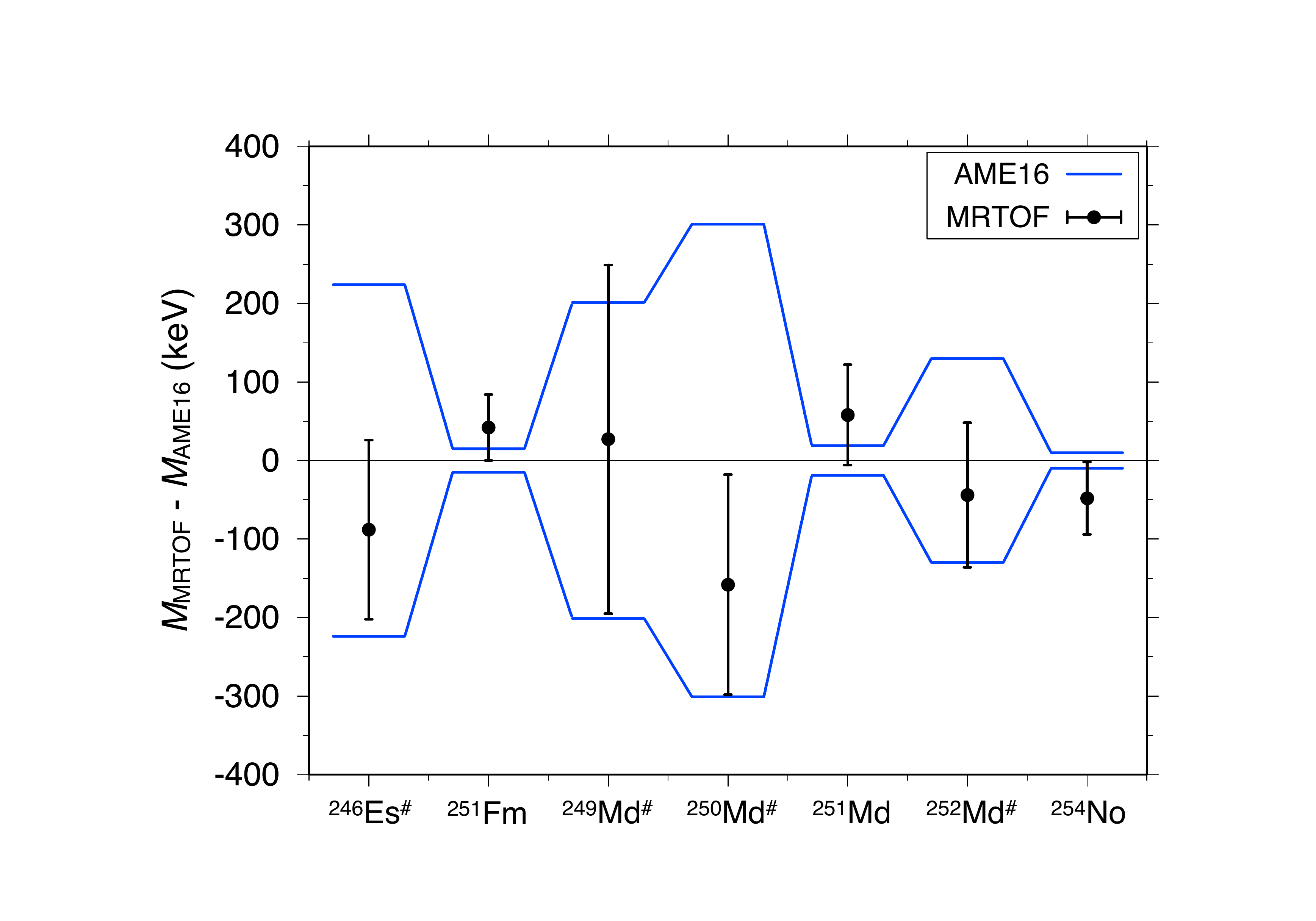}
\caption{Deviations between mass values determined in this work and AME16 \cite{ame2016:2} values. Error bars indicate 1$\sigma$ standard uncertainty of our data, while solid lines indicate uncertainty of AME16. Isotopes designated with superscript \# have extrapolated mass values in AME16.}
\label{fig:mass_dev}
\end{figure}

 %\textcircled{\scriptsize 9}
Results for all isotopes measured are listed in Table~\ref{tab:mass_excess}; the mass values are compared with AME16 \cite{ame2016:2} values in Fig.~\ref{fig:mass_dev}.
The masses were derived using the single-reference analysis method described in \cite{ito2013011306}.
The listed systematic uncertainties derive from ambiguity in the origin of the time-of-flight.
%\textcolor{red}{The systematic uncertainties from possible uncertainty in the origin of time-of-flight involving the single-reference analysis method were evaluated.}
As expected from the short measurement cycle, ToF spectra for $^{254}$No included a $\sim$30\% admixture \cite{hessberger201055} of the 1.295(2)~MeV isomer.  While the isomer and ground state could only be partially resolved, the mass of $^{254\mathrm{g}}$No is consistent with prior direct measurements at the Penning trap mass spectrometer SHIPTRAP \cite{block2010785}. Furthermore, the masses of $^{251}$Fm and $^{251}$Md are in good agreement with those determined by $Q_{\alpha}$ \cite{huand2017030002} using SHIPTRAP values for the masses of $^{255}$Lr and $^{255}$No; this work provides the first direct mass measurements of $^{251}$Fm and $^{251}$Md.  In the cases of $^{246}$Es and $^{249,250,252}$Md, no previous experimental mass data exist, however our values are consistent with extrapolated mass values in AME16 with similar or higher mass precisions.

\par
\begin{figure}[t]
\includegraphics[width=\linewidth]{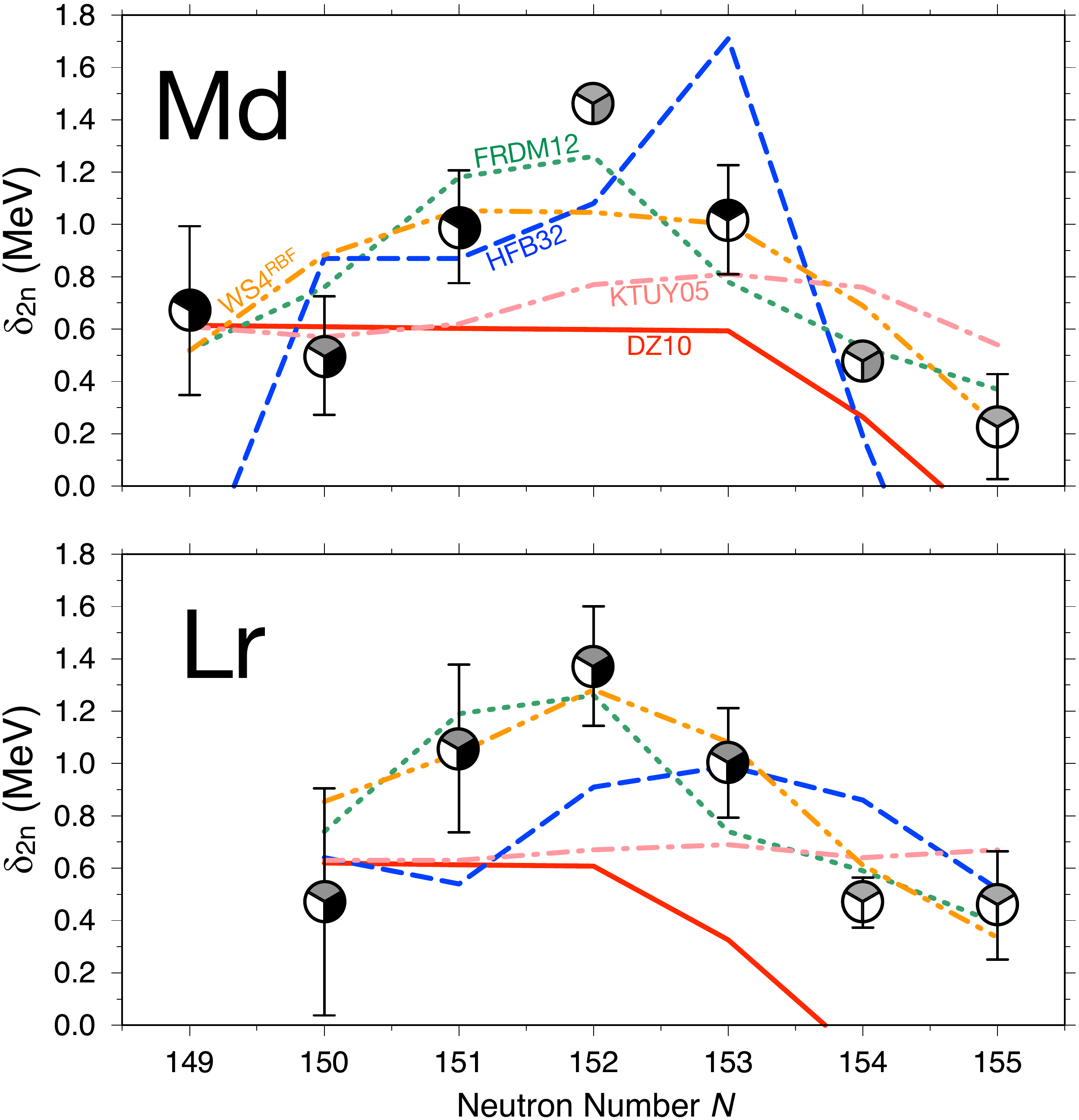}
\caption{Plots of empirical shell gap $\delta_{2n}$ for Md and Lr isotopes. Data points are divided into three sections to indicate the contributions of the three mass values used in each.  White indicates extrapolated value from AME16, gray indicates experimental value from AME16, and black indicates experimental values from this work.  Lines indicate results of theoretical models.}
\label{fig:d2n_Md-Lr}
\end{figure}

 %\textcircled{\scriptsize 10}
One important test applied to theoretical models is their ability to reproduce the shell gap parameter $\delta_{2n}$ \cite{rutz1997238}.
The shell gap parameter $\delta_{2n}$ is calculated as
\begin{eqnarray}
\delta_{2n}(N,Z) &=& S_{2n}(N,Z) - S_{2n}(N+2,Z) \nonumber \\
                 &=& 2B(N,Z) - B(N-2,Z) \nonumber \\
                 & &-B(N+2,Z),
\label{eq:d2n}
\end{eqnarray}
where $S_{2n}(N,Z)$ and $B(N,Z)$ are the two neutron separation energy and the total binding energy of nuclide $^{N+Z}Z$.  Newly determined $\delta_{2n}(N,Z)$ values around $N=152$ for mendelevium and lawrencium are compared with the theoretical values in Fig.~\ref{fig:d2n_Md-Lr}.  As described in Eq.~(\ref{eq:d2n}), three isotopic masses -- $M(N,Z)$, $M(N-2,Z)$, and $M(N+2,Z)$ -- are necessary to derive $\delta_{2n}$.  Using the new mass data in this work, we can make such a comparison between experimental and theoretical values for Md and Lr in the vicinity of the $N=152$ subshell closure.  For comparison to theory, we have selected global mass models representative of various common theoretical techniques: a shell model (DZ10 \cite{duflo1995R23}), macroscopic-microscopic model (FRDM12 \cite{moller20161}, and WS4$^{\mathrm{RBF}}$ \cite{wang2011051303}), a self-consistent mean-field model (HFB32 \cite{goriely2016034337}) and a phenomenological mass model (KTUY05 \cite{koura2005305}).  Both macroscopic-microscopic mass models (FRDM12 and WS4$^{\mathrm{RBF}}$) reasonably predict the experimentally determined $\delta_{2n}$ trends, although the peaking at $N=152$ is best reproduced by the FRDM12 model.  HFB32 and KTUY05 peak beyond $N=152$, while DZ10 shows a flat trend with no peak.
For lawrencium, WS4$^{\mathrm{RBF}}$ agrees well with both the general trend and the peak at $N=152$.

\begin{table}[tb]
\caption{Indirect determination of mass excess ($ME$) using $\alpha$-decay $Q$-values ($Q_{\alpha}$) taken from AME16 \cite{huand2017030002}, along with the AME16 mass excesses ($ME_\mathrm{AME16}$) and our deviation ($\Delta m$) from them.}%
\label{tab:me_002}
\begin{ruledtabular}
\begin{tabular}{cccccccccccccc}
Isotope & $Q_{\alpha}$ & $ME^{\rm Indirect}_{\rm MRTOF}$ & $ME_{\rm AME16}$ & $\Delta m$ \\
 & (keV/$c^{2}$) & (keV/$c^{2}$) & (keV/$c^{2}$) & (keV) \\\hline
$^{253}$Lr &  8918(20) &  88602(222) &  88575$^{\#}$(202$^{\#}$) &   27(222) \\
$^{254}$Lr &  8816(12) &  89713(141) &  89871$^{\#}$(301$^{\#}$) & -158(141) \\
$^{257}$Db &  9207(20) & 100234(224) & 100206$^{\#}$(203$^{\#}$) &   28(224) \\
$^{258}$Db &  9500(50) & 101638(149) & 101797$^{\#}$(306$^{\#}$) & -159(149) \\
$^{261}$Bh & 10500(50) & 113158(229) & 113134$^{\#}$(209$^{\#}$) &   25(229) \\
$^{262}$Bh & 10319(15) & 114382(150) & 114541$^{\#}$(306$^{\#}$) & -159(150) \\
$^{266}$Mt & 10996(25) & 127803(152) & 127962$^{\#}$(306$^{\#}$) & -159(152)
\end{tabular}
\end{ruledtabular}
\end{table}

\par
\begin{figure}[h]
\includegraphics[width=\linewidth]{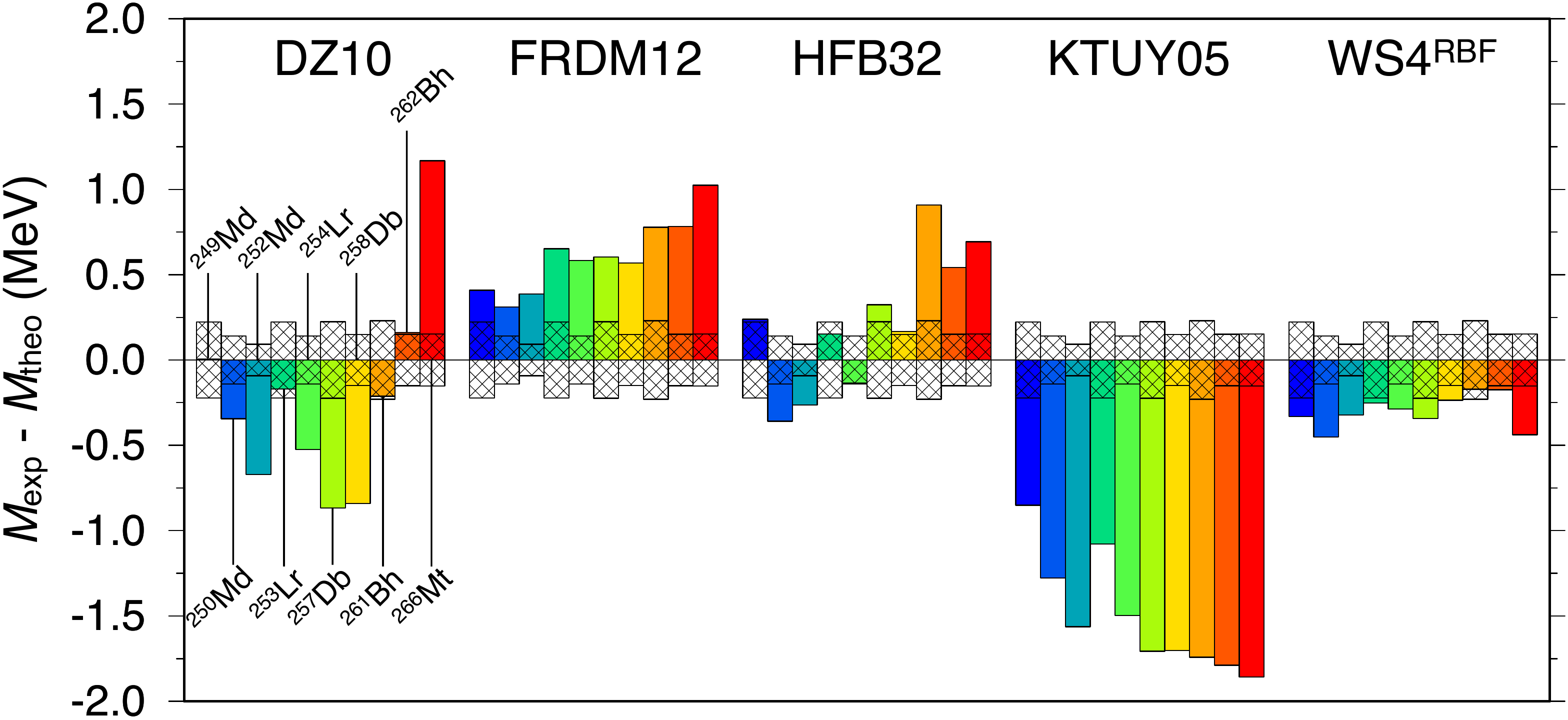}
\caption{Comparison of experimental masses with mass models. Each bar in a model corresponds to an isotope whose mass was determined for the first time in this work: $^{249,250,252}$Md, $^{253,254}$Lr, $^{257,258}$Db, $^{261,262}$Bh, and $^{266}$Mt. Meshed bars indicate experimentally determined mass uncertainties.}
\label{fig:ME_exp-model}
\end{figure}

 %\textcircled{\scriptsize 11}
The location of the island of stability remains ambiguous.
While experimental mass measurements of nuclei located within the hot-fusion superheavy island including the next deformed shell-closure predicted at $N=162$ would be particularly valuable for this, in general more experimentally determined masses in the trans-uranium region will allow for improved extrapolation of mass values into the presumed region of the island of stability.
By supplementing our precision, direct mass measurements with $\alpha$-decay $Q$-values, we can provide mass values for nuclei up to $^{266}$Mt, as listed in Table~\ref{tab:me_002}.  For future efforts to approach to the island of stability, reliable theoretical predictions are crucial.
%\textcolor{red}{Particularly, mass prediction powers for unknown nuclei further away from known nuclei are indispensable since the experimentally accessible region nuclei are still limited}.
Figure~\ref{fig:ME_exp-model} shows the deviations between the various theoretical models and our experimental values.  The best average agreement, with a mean deviation below 500 keV/c$^2$, is obtained from the WS4$^\mathrm{RBF}$  mass model, which is based on the WS4 mass model \cite{wang2014215} using a radial basis function approach for prediction. Except for the case of the KTUY05 model, general agreement within about 1 MeV/c$^2$ is observed, although the deviations tend to increase with the mass.

%\section{summary}

\par %\textcircled{\scriptsize 12}
In this study, we have directly measured the masses of $^{246}$Es, $^{251}$Fm, $^{249-252}$Md, and $^{254\mathrm{g,m1}}$No.  In the cases of $^{254\mathrm{g}}$No, $^{251}$Fm, and $^{251}$Md, the AME16 mass values were derived from Penning trap data; the excellent agreement of our measurement with these Penning trap derived data provides a high degree of confidence in our experimental technique.  Combining these results with $\alpha$-decay $Q$ values, the masses of $^{253,254}$Lr, $^{257, 258}$Db, $^{261,262}$Bh, and $^{266}$Mt could be indirectly determined.  Particularly robust agreement is seen with the WS4$^\mathrm{RBF}$ mass model.

\par %\textcircled{\scriptsize 13}
This work demonstrates the ability to perform direct mass measurements of both cold- and hot-fusion products, even with low recoil energy products ($E_\mathrm{recoil}\approx7$~MeV), by coupling a gas cell with GARIS-II. This technique could be applied to most nuclei produced with fusion-evaporation reactions in the SHN region.
The overall system efficiency behind GARIS-II, excluding GARIS-II efficiency, from stopping in the gas cell to detection, was $\sim$2\% limited by the double trap system.
%, i.e.\ transport and re-trapping efficiencies.
In the near future, modification to a single trap setup at a new experimental location should provide improved system efficiency of more than 10\% and shorter measurement time. 
This will allow us to measure the masses of hot-fusion SHN having cross sections on the order of ten picobarn.
%, to provide clear identification of the atomic and mass numbers.

\par %\textcircled{\scriptsize 14}
We wish to express gratitude to the Nishina Center for Accelerator-based Research at RIKEN and the Center for Nuclear Science at the University of Tokyo for their support during the online experiments. This work was financially supported by the Japan Society for the Promotion of Science KAKENHI (Grant Nos. 2200823, 24224008, 24740142, 15H02096, 15K05116, and 17H06090).

\bibliography{ito_Md_paper_v08_PRL_proof_v02}

\providecommand{\noopsort}[1]{}\providecommand{\singleletter}[1]{#1}%
\begin{thebibliography}{41}%
\makeatletter
\providecommand \@ifxundefined [1]{%
 \@ifx{#1\undefined}
}%
\providecommand \@ifnum [1]{%
 \ifnum #1\expandafter \@firstoftwo
 \else \expandafter \@secondoftwo
 \fi
}%
\providecommand \@ifx [1]{%
 \ifx #1\expandafter \@firstoftwo
 \else \expandafter \@secondoftwo
 \fi
}%
\providecommand \natexlab [1]{#1}%
\providecommand \enquote  [1]{``#1''}%
\providecommand \bibnamefont  [1]{#1}%
\providecommand \bibfnamefont [1]{#1}%
\providecommand \citenamefont [1]{#1}%
\providecommand \href@noop [0]{\@secondoftwo}%
\providecommand \href [0]{\begingroup \@sanitize@url \@href}%
\providecommand \@href[1]{\@@startlink{#1}\@@href}%
\providecommand \@@href[1]{\endgroup#1\@@endlink}%
\providecommand \@sanitize@url [0]{\catcode `\\12\catcode `\$12\catcode
  `\&12\catcode `\#12\catcode `\^12\catcode `\_12\catcode `\%12\relax}%
\providecommand \@@startlink[1]{}%
\providecommand \@@endlink[0]{}%
\providecommand \url  [0]{\begingroup\@sanitize@url \@url }%
\providecommand \@url [1]{\endgroup\@href {#1}{\urlprefix }}%
\providecommand \urlprefix  [0]{URL }%
\providecommand \Eprint [0]{\href }%
\providecommand \doibase [0]{http://dx.doi.org/}%
\providecommand \selectlanguage [0]{\@gobble}%
\providecommand \bibinfo  [0]{\@secondoftwo}%
\providecommand \bibfield  [0]{\@secondoftwo}%
\providecommand \translation [1]{[#1]}%
\providecommand \BibitemOpen [0]{}%
\providecommand \bibitemStop [0]{}%
\providecommand \bibitemNoStop [0]{.\EOS\space}%
\providecommand \EOS [0]{\spacefactor3000\relax}%
\providecommand \BibitemShut  [1]{\csname bibitem#1\endcsname}%
\let\auto@bib@innerbib\@empty
%</preamble>
\bibitem [{\citenamefont {Lunney}\ \emph {et~al.}(2003)\citenamefont {Lunney},
  \citenamefont {Pearson},\ and\ \citenamefont {Thibault}}]{Lunney:2003ue}%
  \BibitemOpen
  \bibfield  {author} {\bibinfo {author} {\bibfnamefont {D.}~\bibnamefont
  {Lunney}}, \bibinfo {author} {\bibfnamefont {J.~M.}\ \bibnamefont {Pearson}},
  \ and\ \bibinfo {author} {\bibfnamefont {C.}~\bibnamefont {Thibault}},\
  }\href@noop {} {\bibfield  {journal} {\bibinfo  {journal} {Reviews Of Modern
  Physics}\ }\textbf {\bibinfo {volume} {75}},\ \bibinfo {pages} {1021}
  (\bibinfo {year} {2003})}\BibitemShut {NoStop}%
\bibitem [{\citenamefont {Sorlin}\ and\ \citenamefont
  {Porquet}(2008)}]{SORLIN2008a}%
  \BibitemOpen
  \bibfield  {author} {\bibinfo {author} {\bibfnamefont {O.}~\bibnamefont
  {Sorlin}}\ and\ \bibinfo {author} {\bibfnamefont {M.-G.}\ \bibnamefont
  {Porquet}},\ }\href {\doibase http://dx.doi.org/10.1016/j.ppnp.2008.05.001}
  {\bibfield  {journal} {\bibinfo  {journal} {Progress in Particle and Nuclear
  Physics}\ }\textbf {\bibinfo {volume} {61}},\ \bibinfo {pages} {602 }
  (\bibinfo {year} {2008})}\BibitemShut {NoStop}%
\bibitem [{\citenamefont {Sobiczewski}\ \emph {et~al.}(1966)\citenamefont
  {Sobiczewski}, \citenamefont {Gareev},\ and\ \citenamefont
  {Kalinkin}}]{SOBICZEWSKI1966500}%
  \BibitemOpen
  \bibfield  {author} {\bibinfo {author} {\bibfnamefont {A.}~\bibnamefont
  {Sobiczewski}}, \bibinfo {author} {\bibfnamefont {F.}~\bibnamefont {Gareev}},
  \ and\ \bibinfo {author} {\bibfnamefont {B.}~\bibnamefont {Kalinkin}},\
  }\href {\doibase http://dx.doi.org/10.1016/0031-9163(66)91243-1} {\bibfield
  {journal} {\bibinfo  {journal} {Physics Letters}\ }\textbf {\bibinfo {volume}
  {22}},\ \bibinfo {pages} {500 } (\bibinfo {year} {1966})}\BibitemShut
  {NoStop}%
\bibitem [{\citenamefont {Patyk}\ \emph {et~al.}(1989)\citenamefont {Patyk},
  \citenamefont {Sobiczewski}, \citenamefont {Armbruster},\ and\ \citenamefont
  {Schmidt}}]{PATYK1989267}%
  \BibitemOpen
  \bibfield  {author} {\bibinfo {author} {\bibfnamefont {Z.}~\bibnamefont
  {Patyk}}, \bibinfo {author} {\bibfnamefont {A.}~\bibnamefont {Sobiczewski}},
  \bibinfo {author} {\bibfnamefont {P.}~\bibnamefont {Armbruster}}, \ and\
  \bibinfo {author} {\bibfnamefont {K.-H.}\ \bibnamefont {Schmidt}},\ }\href
  {\doibase http://dx.doi.org/10.1016/0375-9474(89)90702-1} {\bibfield
  {journal} {\bibinfo  {journal} {Nuclear Physics A}\ }\textbf {\bibinfo
  {volume} {491}},\ \bibinfo {pages} {267 } (\bibinfo {year}
  {1989})}\BibitemShut {NoStop}%
\bibitem [{\citenamefont {Patyk}\ and\ \citenamefont
  {Sobiczewski}(1991)}]{PATYK1991132}%
  \BibitemOpen
  \bibfield  {author} {\bibinfo {author} {\bibfnamefont {Z.}~\bibnamefont
  {Patyk}}\ and\ \bibinfo {author} {\bibfnamefont {A.}~\bibnamefont
  {Sobiczewski}},\ }\href {\doibase
  http://dx.doi.org/10.1016/0375-9474(91)90823-O} {\bibfield  {journal}
  {\bibinfo  {journal} {Nuclear Physics A}\ }\textbf {\bibinfo {volume}
  {533}},\ \bibinfo {pages} {132 } (\bibinfo {year} {1991})}\BibitemShut
  {NoStop}%
\bibitem [{\citenamefont {Seaborg}(1987)}]{seaborg1987}%
  \BibitemOpen
  \bibfield  {author} {\bibinfo {author} {\bibfnamefont {G.~I.}\ \bibnamefont
  {Seaborg}},\ }\href@noop {} {\bibfield  {journal} {\bibinfo  {journal}
  {Contemporary Physics}\ }\textbf {\bibinfo {volume} {28}},\ \bibinfo {pages}
  {33} (\bibinfo {year} {1987})}\BibitemShut {NoStop}%
\bibitem [{\citenamefont {Bender}\ \emph {et~al.}(2001)\citenamefont {Bender},
  \citenamefont {Nazarewicz},\ and\ \citenamefont {Reinhard}}]{BENDER200142}%
  \BibitemOpen
  \bibfield  {author} {\bibinfo {author} {\bibfnamefont {M.}~\bibnamefont
  {Bender}}, \bibinfo {author} {\bibfnamefont {W.}~\bibnamefont {Nazarewicz}},
  \ and\ \bibinfo {author} {\bibfnamefont {P.-G.}\ \bibnamefont {Reinhard}},\
  }\href {\doibase http://dx.doi.org/10.1016/S0370-2693(01)00863-2} {\bibfield
  {journal} {\bibinfo  {journal} {Physics Letters B}\ }\textbf {\bibinfo
  {volume} {515}},\ \bibinfo {pages} {42 } (\bibinfo {year}
  {2001})}\BibitemShut {NoStop}%
\bibitem [{\citenamefont {Cwiok}\ \emph {et~al.}(2005)\citenamefont {Cwiok},
  \citenamefont {Heenen},\ and\ \citenamefont {Nazarewicz}}]{Cwiok2005}%
  \BibitemOpen
  \bibfield  {author} {\bibinfo {author} {\bibfnamefont {S.}~\bibnamefont
  {Cwiok}}, \bibinfo {author} {\bibfnamefont {P.-H.}\ \bibnamefont {Heenen}}, \
  and\ \bibinfo {author} {\bibfnamefont {W.}~\bibnamefont {Nazarewicz}},\
  }\href {\doibase 10.1038/nature03336} {\bibfield  {journal} {\bibinfo
  {journal} {Nature}\ }\textbf {\bibinfo {volume} {433}},\ \bibinfo {pages}
  {705} (\bibinfo {year} {2005})}\BibitemShut {NoStop}%
\bibitem [{\citenamefont {Lazarev}\ \emph {et~al.}(1996)\citenamefont
  {Lazarev}, \citenamefont {Lobanov}, \citenamefont {Oganessian}, \citenamefont
  {Utyonkov}, \citenamefont {Abdullin}, \citenamefont {Polyakov}, \citenamefont
  {Rigol}, \citenamefont {Shirokovsky}, \citenamefont {Tsyganov}, \citenamefont
  {Iliev}, \citenamefont {Subbotin}, \citenamefont {Sukhov}, \citenamefont
  {Buklanov}, \citenamefont {Gikal}, \citenamefont {Kutner}, \citenamefont
  {Mezentsev}, \citenamefont {Subotic}, \citenamefont {Wild}, \citenamefont
  {Lougheed},\ and\ \citenamefont {Moody}}]{PhysRevC.54.620}%
  \BibitemOpen
  \bibfield  {author} {\bibinfo {author} {\bibfnamefont {Y.~A.}\ \bibnamefont
  {Lazarev}}, \bibinfo {author} {\bibfnamefont {Y.~V.}\ \bibnamefont
  {Lobanov}}, \bibinfo {author} {\bibfnamefont {Y.~T.}\ \bibnamefont
  {Oganessian}}, \bibinfo {author} {\bibfnamefont {V.~K.}\ \bibnamefont
  {Utyonkov}}, \bibinfo {author} {\bibfnamefont {F.~S.}\ \bibnamefont
  {Abdullin}}, \bibinfo {author} {\bibfnamefont {A.~N.}\ \bibnamefont
  {Polyakov}}, \bibinfo {author} {\bibfnamefont {J.}~\bibnamefont {Rigol}},
  \bibinfo {author} {\bibfnamefont {I.~V.}\ \bibnamefont {Shirokovsky}},
  \bibinfo {author} {\bibfnamefont {Y.~S.}\ \bibnamefont {Tsyganov}}, \bibinfo
  {author} {\bibfnamefont {S.}~\bibnamefont {Iliev}}, \bibinfo {author}
  {\bibfnamefont {V.~G.}\ \bibnamefont {Subbotin}}, \bibinfo {author}
  {\bibfnamefont {A.~M.}\ \bibnamefont {Sukhov}}, \bibinfo {author}
  {\bibfnamefont {G.~V.}\ \bibnamefont {Buklanov}}, \bibinfo {author}
  {\bibfnamefont {B.~N.}\ \bibnamefont {Gikal}}, \bibinfo {author}
  {\bibfnamefont {V.~B.}\ \bibnamefont {Kutner}}, \bibinfo {author}
  {\bibfnamefont {A.~N.}\ \bibnamefont {Mezentsev}}, \bibinfo {author}
  {\bibfnamefont {K.}~\bibnamefont {Subotic}}, \bibinfo {author} {\bibfnamefont
  {J.~F.}\ \bibnamefont {Wild}}, \bibinfo {author} {\bibfnamefont {R.~W.}\
  \bibnamefont {Lougheed}}, \ and\ \bibinfo {author} {\bibfnamefont {K.~J.}\
  \bibnamefont {Moody}},\ }\href {\doibase 10.1103/PhysRevC.54.620} {\bibfield
  {journal} {\bibinfo  {journal} {Phys. Rev. C}\ }\textbf {\bibinfo {volume}
  {54}},\ \bibinfo {pages} {620} (\bibinfo {year} {1996})}\BibitemShut
  {NoStop}%
\bibitem [{\citenamefont {Oganessian}(2011)}]{Oganessian:2011cm}%
  \BibitemOpen
  \bibfield  {author} {\bibinfo {author} {\bibfnamefont {Y.~T.}\ \bibnamefont
  {Oganessian}},\ }\href@noop {} {\bibfield  {journal} {\bibinfo  {journal}
  {Radiochimica Acta}\ }\textbf {\bibinfo {volume} {99}},\ \bibinfo {pages}
  {429} (\bibinfo {year} {2011})}\BibitemShut {NoStop}%
\bibitem [{\citenamefont {Rudolph}\ \emph {et~al.}(2013)\citenamefont
  {Rudolph}, \citenamefont {Forsberg}, \citenamefont {Golubev}, \citenamefont
  {Sarmiento}, \citenamefont {Yakushev}, \citenamefont {Andersson},
  \citenamefont {Di~Nitto}, \citenamefont {D\"ullmann}, \citenamefont {Gates},
  \citenamefont {Gregorich}, \citenamefont {Gross}, \citenamefont
  {He\ss{}berger}, \citenamefont {Herzberg}, \citenamefont {Khuyagbaatar},
  \citenamefont {Kratz}, \citenamefont {Rykaczewski}, \citenamefont
  {Sch\"adel}, \citenamefont {\AA{}berg}, \citenamefont {Ackermann},
  \citenamefont {Block}, \citenamefont {Brand}, \citenamefont {Carlsson},
  \citenamefont {Cox}, \citenamefont {Derkx}, \citenamefont {Eberhardt},
  \citenamefont {Even}, \citenamefont {Fahlander}, \citenamefont {Gerl},
  \citenamefont {J\"ager}, \citenamefont {Kindler}, \citenamefont {Krier},
  \citenamefont {Kojouharov}, \citenamefont {Kurz}, \citenamefont {Lommel},
  \citenamefont {Mistry}, \citenamefont {Mokry}, \citenamefont {Nitsche},
  \citenamefont {Omtvedt}, \citenamefont {Papadakis}, \citenamefont
  {Ragnarsson}, \citenamefont {Runke}, \citenamefont {Schaffner}, \citenamefont
  {Schausten}, \citenamefont {Th\"orle-Pospiech}, \citenamefont {Torres},
  \citenamefont {Traut}, \citenamefont {Trautmann}, \citenamefont {T\"urler},
  \citenamefont {Ward}, \citenamefont {Ward},\ and\ \citenamefont
  {Wiehl}}]{Rudolph2013}%
  \BibitemOpen
  \bibfield  {author} {\bibinfo {author} {\bibfnamefont {D.}~\bibnamefont
  {Rudolph}}, \bibinfo {author} {\bibfnamefont {U.}~\bibnamefont {Forsberg}},
  \bibinfo {author} {\bibfnamefont {P.}~\bibnamefont {Golubev}}, \bibinfo
  {author} {\bibfnamefont {L.~G.}\ \bibnamefont {Sarmiento}}, \bibinfo {author}
  {\bibfnamefont {A.}~\bibnamefont {Yakushev}}, \bibinfo {author}
  {\bibfnamefont {L.-L.}\ \bibnamefont {Andersson}}, \bibinfo {author}
  {\bibfnamefont {A.}~\bibnamefont {Di~Nitto}}, \bibinfo {author}
  {\bibfnamefont {C.~E.}\ \bibnamefont {D\"ullmann}}, \bibinfo {author}
  {\bibfnamefont {J.~M.}\ \bibnamefont {Gates}}, \bibinfo {author}
  {\bibfnamefont {K.~E.}\ \bibnamefont {Gregorich}}, \bibinfo {author}
  {\bibfnamefont {C.~J.}\ \bibnamefont {Gross}}, \bibinfo {author}
  {\bibfnamefont {F.~P.}\ \bibnamefont {He\ss{}berger}}, \bibinfo {author}
  {\bibfnamefont {R.-D.}\ \bibnamefont {Herzberg}}, \bibinfo {author}
  {\bibfnamefont {J.}~\bibnamefont {Khuyagbaatar}}, \bibinfo {author}
  {\bibfnamefont {J.~V.}\ \bibnamefont {Kratz}}, \bibinfo {author}
  {\bibfnamefont {K.}~\bibnamefont {Rykaczewski}}, \bibinfo {author}
  {\bibfnamefont {M.}~\bibnamefont {Sch\"adel}}, \bibinfo {author}
  {\bibfnamefont {S.}~\bibnamefont {\AA{}berg}}, \bibinfo {author}
  {\bibfnamefont {D.}~\bibnamefont {Ackermann}}, \bibinfo {author}
  {\bibfnamefont {M.}~\bibnamefont {Block}}, \bibinfo {author} {\bibfnamefont
  {H.}~\bibnamefont {Brand}}, \bibinfo {author} {\bibfnamefont {B.~G.}\
  \bibnamefont {Carlsson}}, \bibinfo {author} {\bibfnamefont {D.}~\bibnamefont
  {Cox}}, \bibinfo {author} {\bibfnamefont {X.}~\bibnamefont {Derkx}}, \bibinfo
  {author} {\bibfnamefont {K.}~\bibnamefont {Eberhardt}}, \bibinfo {author}
  {\bibfnamefont {J.}~\bibnamefont {Even}}, \bibinfo {author} {\bibfnamefont
  {C.}~\bibnamefont {Fahlander}}, \bibinfo {author} {\bibfnamefont
  {J.}~\bibnamefont {Gerl}}, \bibinfo {author} {\bibfnamefont {E.}~\bibnamefont
  {J\"ager}}, \bibinfo {author} {\bibfnamefont {B.}~\bibnamefont {Kindler}},
  \bibinfo {author} {\bibfnamefont {J.}~\bibnamefont {Krier}}, \bibinfo
  {author} {\bibfnamefont {I.}~\bibnamefont {Kojouharov}}, \bibinfo {author}
  {\bibfnamefont {N.}~\bibnamefont {Kurz}}, \bibinfo {author} {\bibfnamefont
  {B.}~\bibnamefont {Lommel}}, \bibinfo {author} {\bibfnamefont
  {A.}~\bibnamefont {Mistry}}, \bibinfo {author} {\bibfnamefont
  {C.}~\bibnamefont {Mokry}}, \bibinfo {author} {\bibfnamefont
  {H.}~\bibnamefont {Nitsche}}, \bibinfo {author} {\bibfnamefont {J.~P.}\
  \bibnamefont {Omtvedt}}, \bibinfo {author} {\bibfnamefont {P.}~\bibnamefont
  {Papadakis}}, \bibinfo {author} {\bibfnamefont {I.}~\bibnamefont
  {Ragnarsson}}, \bibinfo {author} {\bibfnamefont {J.}~\bibnamefont {Runke}},
  \bibinfo {author} {\bibfnamefont {H.}~\bibnamefont {Schaffner}}, \bibinfo
  {author} {\bibfnamefont {B.}~\bibnamefont {Schausten}}, \bibinfo {author}
  {\bibfnamefont {P.}~\bibnamefont {Th\"orle-Pospiech}}, \bibinfo {author}
  {\bibfnamefont {T.}~\bibnamefont {Torres}}, \bibinfo {author} {\bibfnamefont
  {T.}~\bibnamefont {Traut}}, \bibinfo {author} {\bibfnamefont
  {N.}~\bibnamefont {Trautmann}}, \bibinfo {author} {\bibfnamefont
  {A.}~\bibnamefont {T\"urler}}, \bibinfo {author} {\bibfnamefont
  {A.}~\bibnamefont {Ward}}, \bibinfo {author} {\bibfnamefont {D.~E.}\
  \bibnamefont {Ward}}, \ and\ \bibinfo {author} {\bibfnamefont
  {N.}~\bibnamefont {Wiehl}},\ }\href {\doibase 10.1103/PhysRevLett.111.112502}
  {\bibfield  {journal} {\bibinfo  {journal} {Phys. Rev. Lett.}\ }\textbf
  {\bibinfo {volume} {111}},\ \bibinfo {pages} {112502} (\bibinfo {year}
  {2013})}\BibitemShut {NoStop}%
\bibitem [{\citenamefont {Julin}(2001)}]{JULIN2001221}%
  \BibitemOpen
  \bibfield  {author} {\bibinfo {author} {\bibfnamefont {R.}~\bibnamefont
  {Julin}},\ }\href {\doibase http://dx.doi.org/10.1016/S0375-9474(01)00542-5}
  {\bibfield  {journal} {\bibinfo  {journal} {Nuclear Physics A}\ }\textbf
  {\bibinfo {volume} {685}},\ \bibinfo {pages} {221 } (\bibinfo {year}
  {2001})},\ \bibinfo {note} {nucleus-Nucleus Collisions 2000}\BibitemShut
  {NoStop}%
\bibitem [{\citenamefont {Bucurescu}\ and\ \citenamefont
  {Zamfir}(2013)}]{PhysRevC.87.054324}%
  \BibitemOpen
  \bibfield  {author} {\bibinfo {author} {\bibfnamefont {D.}~\bibnamefont
  {Bucurescu}}\ and\ \bibinfo {author} {\bibfnamefont {N.~V.}\ \bibnamefont
  {Zamfir}},\ }\href {\doibase 10.1103/PhysRevC.87.054324} {\bibfield
  {journal} {\bibinfo  {journal} {Phys. Rev. C}\ }\textbf {\bibinfo {volume}
  {87}},\ \bibinfo {pages} {054324} (\bibinfo {year} {2013})}\BibitemShut
  {NoStop}%
\bibitem [{\citenamefont {Block}\ \emph {et~al.}(2010)\citenamefont {Block},
  \citenamefont {Ackermann}, \citenamefont {Blaum}, \citenamefont {Droese},
  \citenamefont {Dworschak}, \citenamefont {Eliseev}, \citenamefont
  {Fleckenstein}, \citenamefont {Haettner}, \citenamefont {Herfurth},
  \citenamefont {He{\ss}berger}, \citenamefont {Hofmann}, \citenamefont
  {Ketelaer}, \citenamefont {Ketter}, \citenamefont {Kluge}, \citenamefont
  {Marx}, \citenamefont {Mazzocco}, \citenamefont {Novikov}, \citenamefont
  {Pla{\ss}}, \citenamefont {Popeko}, \citenamefont {Rahaman}, \citenamefont
  {Rodr\'{\i}guez}, \citenamefont {Scheidenberger}, \citenamefont
  {Schweikhard}, \citenamefont {Thirolf}, \citenamefont {Vorobyev},\ and\
  \citenamefont {Weber}}]{block2010785}%
  \BibitemOpen
  \bibfield  {author} {\bibinfo {author} {\bibfnamefont {M.}~\bibnamefont
  {Block}}, \bibinfo {author} {\bibfnamefont {D.}~\bibnamefont {Ackermann}},
  \bibinfo {author} {\bibfnamefont {K.}~\bibnamefont {Blaum}}, \bibinfo
  {author} {\bibfnamefont {C.}~\bibnamefont {Droese}}, \bibinfo {author}
  {\bibfnamefont {M.}~\bibnamefont {Dworschak}}, \bibinfo {author}
  {\bibfnamefont {S.}~\bibnamefont {Eliseev}}, \bibinfo {author} {\bibfnamefont
  {T.}~\bibnamefont {Fleckenstein}}, \bibinfo {author} {\bibfnamefont
  {E.}~\bibnamefont {Haettner}}, \bibinfo {author} {\bibfnamefont
  {F.}~\bibnamefont {Herfurth}}, \bibinfo {author} {\bibfnamefont {F.~P.}\
  \bibnamefont {He{\ss}berger}}, \bibinfo {author} {\bibfnamefont
  {S.}~\bibnamefont {Hofmann}}, \bibinfo {author} {\bibfnamefont
  {J.}~\bibnamefont {Ketelaer}}, \bibinfo {author} {\bibfnamefont
  {J.}~\bibnamefont {Ketter}}, \bibinfo {author} {\bibfnamefont {H.-J.}\
  \bibnamefont {Kluge}}, \bibinfo {author} {\bibfnamefont {G.}~\bibnamefont
  {Marx}}, \bibinfo {author} {\bibfnamefont {M.}~\bibnamefont {Mazzocco}},
  \bibinfo {author} {\bibfnamefont {Y.~N.}\ \bibnamefont {Novikov}}, \bibinfo
  {author} {\bibfnamefont {W.~R.}\ \bibnamefont {Pla{\ss}}}, \bibinfo {author}
  {\bibfnamefont {A.}~\bibnamefont {Popeko}}, \bibinfo {author} {\bibfnamefont
  {S.}~\bibnamefont {Rahaman}}, \bibinfo {author} {\bibfnamefont
  {D.}~\bibnamefont {Rodr\'{\i}guez}}, \bibinfo {author} {\bibfnamefont
  {C.}~\bibnamefont {Scheidenberger}}, \bibinfo {author} {\bibfnamefont
  {L.}~\bibnamefont {Schweikhard}}, \bibinfo {author} {\bibfnamefont {P.~G.}\
  \bibnamefont {Thirolf}}, \bibinfo {author} {\bibfnamefont {G.~K.}\
  \bibnamefont {Vorobyev}}, \ and\ \bibinfo {author} {\bibfnamefont
  {C.}~\bibnamefont {Weber}},\ }\href@noop {} {\bibfield  {journal} {\bibinfo
  {journal} {Nature}\ }\textbf {\bibinfo {volume} {463}},\ \bibinfo {pages}
  {785} (\bibinfo {year} {2010})}\BibitemShut {NoStop}%
\bibitem [{\citenamefont {Ramirez}\ \emph {et~al.}(2012)\citenamefont
  {Ramirez}, \citenamefont {Ackermann}, \citenamefont {Blaum}, \citenamefont
  {Block}, \citenamefont {Droese}, \citenamefont {D\"ullmann}, \citenamefont
  {Dworschak}, \citenamefont {Eibach}, \citenamefont {Eliseev}, \citenamefont
  {Haettner}, \citenamefont {Herfurth}, \citenamefont {He{\ss}berger},
  \citenamefont {Hofmann}, \citenamefont {Ketelaer}, \citenamefont {Marx},
  \citenamefont {Mazzocco}, \citenamefont {Nesterenko}, \citenamefont
  {Novikov}, \citenamefont {Pla{\ss}}, \citenamefont {Rodr\'{\i}guez},
  \citenamefont {Scheidenberger}, \citenamefont {Schweikhard}, \citenamefont
  {Thirolf},\ and\ \citenamefont {Weber}}]{ramirez20121207}%
  \BibitemOpen
  \bibfield  {author} {\bibinfo {author} {\bibfnamefont {E.~M.}\ \bibnamefont
  {Ramirez}}, \bibinfo {author} {\bibfnamefont {D.}~\bibnamefont {Ackermann}},
  \bibinfo {author} {\bibfnamefont {K.}~\bibnamefont {Blaum}}, \bibinfo
  {author} {\bibfnamefont {M.}~\bibnamefont {Block}}, \bibinfo {author}
  {\bibfnamefont {C.}~\bibnamefont {Droese}}, \bibinfo {author} {\bibfnamefont
  {C.~E.}\ \bibnamefont {D\"ullmann}}, \bibinfo {author} {\bibfnamefont
  {M.}~\bibnamefont {Dworschak}}, \bibinfo {author} {\bibfnamefont
  {M.}~\bibnamefont {Eibach}}, \bibinfo {author} {\bibfnamefont
  {S.}~\bibnamefont {Eliseev}}, \bibinfo {author} {\bibfnamefont
  {E.}~\bibnamefont {Haettner}}, \bibinfo {author} {\bibfnamefont
  {F.}~\bibnamefont {Herfurth}}, \bibinfo {author} {\bibfnamefont {F.~P.}\
  \bibnamefont {He{\ss}berger}}, \bibinfo {author} {\bibfnamefont
  {S.}~\bibnamefont {Hofmann}}, \bibinfo {author} {\bibfnamefont
  {J.}~\bibnamefont {Ketelaer}}, \bibinfo {author} {\bibfnamefont
  {G.}~\bibnamefont {Marx}}, \bibinfo {author} {\bibfnamefont {M.}~\bibnamefont
  {Mazzocco}}, \bibinfo {author} {\bibfnamefont {D.}~\bibnamefont
  {Nesterenko}}, \bibinfo {author} {\bibfnamefont {Y.~N.}\ \bibnamefont
  {Novikov}}, \bibinfo {author} {\bibfnamefont {W.~R.}\ \bibnamefont
  {Pla{\ss}}}, \bibinfo {author} {\bibfnamefont {D.}~\bibnamefont
  {Rodr\'{\i}guez}}, \bibinfo {author} {\bibfnamefont {C.}~\bibnamefont
  {Scheidenberger}}, \bibinfo {author} {\bibfnamefont {L.}~\bibnamefont
  {Schweikhard}}, \bibinfo {author} {\bibfnamefont {P.~G.}\ \bibnamefont
  {Thirolf}}, \ and\ \bibinfo {author} {\bibfnamefont {C.}~\bibnamefont
  {Weber}},\ }\href@noop {} {\bibfield  {journal} {\bibinfo  {journal}
  {Science}\ }\textbf {\bibinfo {volume} {337}},\ \bibinfo {pages} {1207}
  (\bibinfo {year} {2012})}\BibitemShut {NoStop}%
\bibitem [{\citenamefont {Ito}\ \emph {et~al.}(2013{\natexlab{a}})\citenamefont
  {Ito}, \citenamefont {Schury}, \citenamefont {Wada}, \citenamefont {Naimi},
  \citenamefont {Sonoda}, \citenamefont {Mita}, \citenamefont {Arai},
  \citenamefont {Takamine}, \citenamefont {Okada}, \citenamefont {Ozawa},\ and\
  \citenamefont {Wollnik}}]{ito2013011306}%
  \BibitemOpen
  \bibfield  {author} {\bibinfo {author} {\bibfnamefont {Y.}~\bibnamefont
  {Ito}}, \bibinfo {author} {\bibfnamefont {P.}~\bibnamefont {Schury}},
  \bibinfo {author} {\bibfnamefont {M.}~\bibnamefont {Wada}}, \bibinfo {author}
  {\bibfnamefont {S.}~\bibnamefont {Naimi}}, \bibinfo {author} {\bibfnamefont
  {T.}~\bibnamefont {Sonoda}}, \bibinfo {author} {\bibfnamefont
  {H.}~\bibnamefont {Mita}}, \bibinfo {author} {\bibfnamefont {F.}~\bibnamefont
  {Arai}}, \bibinfo {author} {\bibfnamefont {A.}~\bibnamefont {Takamine}},
  \bibinfo {author} {\bibfnamefont {K.}~\bibnamefont {Okada}}, \bibinfo
  {author} {\bibfnamefont {A.}~\bibnamefont {Ozawa}}, \ and\ \bibinfo {author}
  {\bibfnamefont {H.}~\bibnamefont {Wollnik}},\ }\href@noop {} {\bibfield
  {journal} {\bibinfo  {journal} {Phys.\ Rev.\ C}\ }\textbf {\bibinfo {volume}
  {88}},\ \bibinfo {pages} {011306} (\bibinfo {year}
  {2013}{\natexlab{a}})}\BibitemShut {NoStop}%
\bibitem [{\citenamefont {Schury}\ \emph {et~al.}(2014)\citenamefont {Schury},
  \citenamefont {Wada}, \citenamefont {Ito}, \citenamefont {Arai},
  \citenamefont {Naimi}, \citenamefont {Sonoda}, \citenamefont {Wollnik},
  \citenamefont {Shchepunov}, \citenamefont {Smorra},\ and\ \citenamefont
  {Yuan}}]{schury2014nim}%
  \BibitemOpen
  \bibfield  {author} {\bibinfo {author} {\bibfnamefont {P.}~\bibnamefont
  {Schury}}, \bibinfo {author} {\bibfnamefont {M.}~\bibnamefont {Wada}},
  \bibinfo {author} {\bibfnamefont {Y.}~\bibnamefont {Ito}}, \bibinfo {author}
  {\bibfnamefont {F.}~\bibnamefont {Arai}}, \bibinfo {author} {\bibfnamefont
  {S.}~\bibnamefont {Naimi}}, \bibinfo {author} {\bibfnamefont
  {T.}~\bibnamefont {Sonoda}}, \bibinfo {author} {\bibfnamefont
  {H.}~\bibnamefont {Wollnik}}, \bibinfo {author} {\bibfnamefont
  {V.}~\bibnamefont {Shchepunov}}, \bibinfo {author} {\bibfnamefont
  {C.}~\bibnamefont {Smorra}}, \ and\ \bibinfo {author} {\bibfnamefont
  {C.}~\bibnamefont {Yuan}},\ }\href {\doibase
  http://dx.doi.org/10.1016/j.nimb.2014.05.016} {\bibfield  {journal} {\bibinfo
   {journal} {Nuclear Instruments and Methods in Physics Research Section B:
  Beam Interactions with Materials and Atoms}\ }\textbf {\bibinfo {volume}
  {335}},\ \bibinfo {pages} {39 } (\bibinfo {year} {2014})}\BibitemShut
  {NoStop}%
\bibitem [{\citenamefont {Kaji}\ \emph {et~al.}(2013)\citenamefont {Kaji},
  \citenamefont {Morimoto}, \citenamefont {Sato}, \citenamefont {Yoneda},\ and\
  \citenamefont {Morita}}]{kaji2013311}%
  \BibitemOpen
  \bibfield  {author} {\bibinfo {author} {\bibfnamefont {D.}~\bibnamefont
  {Kaji}}, \bibinfo {author} {\bibfnamefont {K.}~\bibnamefont {Morimoto}},
  \bibinfo {author} {\bibfnamefont {N.}~\bibnamefont {Sato}}, \bibinfo {author}
  {\bibfnamefont {A.}~\bibnamefont {Yoneda}}, \ and\ \bibinfo {author}
  {\bibfnamefont {K.}~\bibnamefont {Morita}},\ }\href@noop {} {\bibfield
  {journal} {\bibinfo  {journal} {Nucl.\ Instr.\ Meth.\ B}\ }\textbf {\bibinfo
  {volume} {317}},\ \bibinfo {pages} {311} (\bibinfo {year}
  {2013})}\BibitemShut {NoStop}%
\bibitem [{\citenamefont {Wada}\ \emph {et~al.}(2003)\citenamefont {Wada},
  \citenamefont {Ishida}, \citenamefont {Nakamura}, \citenamefont {Yamazaki},
  \citenamefont {Kambara}, \citenamefont {Ohyama}, \citenamefont {Kanai},
  \citenamefont {Kojima}, \citenamefont {Nakai}, \citenamefont {Ohshima},
  \citenamefont {Yoshida}, \citenamefont {Kubo}, \citenamefont {Matsuo},
  \citenamefont {Fukuyama}, \citenamefont {Okada}, \citenamefont {Sonoda},
  \citenamefont {Ohtani}, \citenamefont {Noda}, \citenamefont {Kawakami},\ and\
  \citenamefont {Katayama}}]{wada2003570}%
  \BibitemOpen
  \bibfield  {author} {\bibinfo {author} {\bibfnamefont {M.}~\bibnamefont
  {Wada}}, \bibinfo {author} {\bibfnamefont {Y.}~\bibnamefont {Ishida}},
  \bibinfo {author} {\bibfnamefont {T.}~\bibnamefont {Nakamura}}, \bibinfo
  {author} {\bibfnamefont {Y.}~\bibnamefont {Yamazaki}}, \bibinfo {author}
  {\bibfnamefont {T.}~\bibnamefont {Kambara}}, \bibinfo {author} {\bibfnamefont
  {H.}~\bibnamefont {Ohyama}}, \bibinfo {author} {\bibfnamefont
  {Y.}~\bibnamefont {Kanai}}, \bibinfo {author} {\bibfnamefont {T.~M.}\
  \bibnamefont {Kojima}}, \bibinfo {author} {\bibfnamefont {Y.}~\bibnamefont
  {Nakai}}, \bibinfo {author} {\bibfnamefont {N.}~\bibnamefont {Ohshima}},
  \bibinfo {author} {\bibfnamefont {A.}~\bibnamefont {Yoshida}}, \bibinfo
  {author} {\bibfnamefont {T.}~\bibnamefont {Kubo}}, \bibinfo {author}
  {\bibfnamefont {Y.}~\bibnamefont {Matsuo}}, \bibinfo {author} {\bibfnamefont
  {Y.}~\bibnamefont {Fukuyama}}, \bibinfo {author} {\bibfnamefont
  {K.}~\bibnamefont {Okada}}, \bibinfo {author} {\bibfnamefont
  {T.}~\bibnamefont {Sonoda}}, \bibinfo {author} {\bibfnamefont
  {S.}~\bibnamefont {Ohtani}}, \bibinfo {author} {\bibfnamefont
  {K.}~\bibnamefont {Noda}}, \bibinfo {author} {\bibfnamefont {H.}~\bibnamefont
  {Kawakami}}, \ and\ \bibinfo {author} {\bibfnamefont {I.}~\bibnamefont
  {Katayama}},\ }\href@noop {} {\bibfield  {journal} {\bibinfo  {journal}
  {Nuclear Instruments and Methods in Physics Research Section B: Beam
  Interactions with Materials and Atoms}\ }\textbf {\bibinfo {volume} {204}},\
  \bibinfo {pages} {570} (\bibinfo {year} {2003})}\BibitemShut {NoStop}%
\bibitem [{\citenamefont {Bollen}(2011)}]{bollen2011131}%
  \BibitemOpen
  \bibfield  {author} {\bibinfo {author} {\bibfnamefont {G.}~\bibnamefont
  {Bollen}},\ }\href@noop {} {\bibfield  {journal} {\bibinfo  {journal} {Int.\
  J.\ Mass Spectrom.}\ }\textbf {\bibinfo {volume} {299}},\ \bibinfo {pages}
  {131} (\bibinfo {year} {2011})}\BibitemShut {NoStop}%
\bibitem [{\citenamefont {Arai}\ \emph {et~al.}(2014)\citenamefont {Arai},
  \citenamefont {Ito}, \citenamefont {Wada}, \citenamefont {Schury},
  \citenamefont {Sonoda},\ and\ \citenamefont {Mita}}]{arai201456}%
  \BibitemOpen
  \bibfield  {author} {\bibinfo {author} {\bibfnamefont {F.}~\bibnamefont
  {Arai}}, \bibinfo {author} {\bibfnamefont {Y.}~\bibnamefont {Ito}}, \bibinfo
  {author} {\bibfnamefont {M.}~\bibnamefont {Wada}}, \bibinfo {author}
  {\bibfnamefont {P.}~\bibnamefont {Schury}}, \bibinfo {author} {\bibfnamefont
  {T.}~\bibnamefont {Sonoda}}, \ and\ \bibinfo {author} {\bibfnamefont
  {H.}~\bibnamefont {Mita}},\ }\href@noop {} {\bibfield  {journal} {\bibinfo
  {journal} {Int.\ J.\ Mass Spectrom.}\ }\textbf {\bibinfo {volume} {362}},\
  \bibinfo {pages} {56} (\bibinfo {year} {2014})}\BibitemShut {NoStop}%
\bibitem [{\citenamefont {Arai}\ \emph {et~al.}(2015)\citenamefont {Arai},
  \citenamefont {Ito}, \citenamefont {Katayama}, \citenamefont {Schury},
  \citenamefont {Sonoda}, \citenamefont {Wada},\ and\ \citenamefont
  {Wollnik}}]{arai2015030110}%
  \BibitemOpen
  \bibfield  {author} {\bibinfo {author} {\bibfnamefont {F.}~\bibnamefont
  {Arai}}, \bibinfo {author} {\bibfnamefont {Y.}~\bibnamefont {Ito}}, \bibinfo
  {author} {\bibfnamefont {I.}~\bibnamefont {Katayama}}, \bibinfo {author}
  {\bibfnamefont {P.}~\bibnamefont {Schury}}, \bibinfo {author} {\bibfnamefont
  {T.}~\bibnamefont {Sonoda}}, \bibinfo {author} {\bibfnamefont
  {M.}~\bibnamefont {Wada}}, \ and\ \bibinfo {author} {\bibfnamefont
  {H.}~\bibnamefont {Wollnik}},\ }\href@noop {} {\bibfield  {journal} {\bibinfo
   {journal} {JPS Conf.\ Proc.}\ }\textbf {\bibinfo {volume} {6}},\ \bibinfo
  {pages} {030110} (\bibinfo {year} {2015})}\BibitemShut {NoStop}%
\bibitem [{\citenamefont {Bradbury}\ and\ \citenamefont
  {Nielsen}(1936)}]{bradbury1936388}%
  \BibitemOpen
  \bibfield  {author} {\bibinfo {author} {\bibfnamefont {N.~E.}\ \bibnamefont
  {Bradbury}}\ and\ \bibinfo {author} {\bibfnamefont {R.~A.}\ \bibnamefont
  {Nielsen}},\ }\href@noop {} {\bibfield  {journal} {\bibinfo  {journal} {Phys.
  Rev.}\ }\textbf {\bibinfo {volume} {49}},\ \bibinfo {pages} {388} (\bibinfo
  {year} {1936})}\BibitemShut {NoStop}%
\bibitem [{\citenamefont {Ito}\ \emph {et~al.}(2013{\natexlab{b}})\citenamefont
  {Ito}, \citenamefont {Schury}, \citenamefont {Wada}, \citenamefont {Naimi},
  \citenamefont {Smorra}, \citenamefont {Sonoda}, \citenamefont {Mita},
  \citenamefont {Takamine}, \citenamefont {Okada}, \citenamefont {Ozawa},\ and\
  \citenamefont {Wollnik}}]{ito2013544}%
  \BibitemOpen
  \bibfield  {author} {\bibinfo {author} {\bibfnamefont {Y.}~\bibnamefont
  {Ito}}, \bibinfo {author} {\bibfnamefont {P.}~\bibnamefont {Schury}},
  \bibinfo {author} {\bibfnamefont {M.}~\bibnamefont {Wada}}, \bibinfo {author}
  {\bibfnamefont {S.}~\bibnamefont {Naimi}}, \bibinfo {author} {\bibfnamefont
  {C.}~\bibnamefont {Smorra}}, \bibinfo {author} {\bibfnamefont
  {T.}~\bibnamefont {Sonoda}}, \bibinfo {author} {\bibfnamefont
  {H.}~\bibnamefont {Mita}}, \bibinfo {author} {\bibfnamefont {A.}~\bibnamefont
  {Takamine}}, \bibinfo {author} {\bibfnamefont {K.}~\bibnamefont {Okada}},
  \bibinfo {author} {\bibfnamefont {A.}~\bibnamefont {Ozawa}}, \ and\ \bibinfo
  {author} {\bibfnamefont {H.}~\bibnamefont {Wollnik}},\ }\href@noop {}
  {\bibfield  {journal} {\bibinfo  {journal} {Nuclear Instruments and Methods
  in Physics Research Section B: Beam Interactions with Materials and Atoms}\
  }\textbf {\bibinfo {volume} {317, Part B}},\ \bibinfo {pages} {544} (\bibinfo
  {year} {2013}{\natexlab{b}})}\BibitemShut {NoStop}%
\bibitem [{\citenamefont {Stancik}\ and\ \citenamefont
  {Brauns}(2008)}]{stancik2008}%
  \BibitemOpen
  \bibfield  {author} {\bibinfo {author} {\bibfnamefont {A.~L.}\ \bibnamefont
  {Stancik}}\ and\ \bibinfo {author} {\bibfnamefont {E.~B.}\ \bibnamefont
  {Brauns}},\ }\href@noop {} {\bibfield  {journal} {\bibinfo  {journal}
  {Vibrational Spectroscopy}\ }\textbf {\bibinfo {volume} {47}},\ \bibinfo
  {pages} {66} (\bibinfo {year} {2008})}\BibitemShut {NoStop}%
\bibitem [{\citenamefont {{R Development Core Team}}(2008)}]{r2008}%
  \BibitemOpen
  \bibfield  {author} {\bibinfo {author} {\bibnamefont {{R Development Core
  Team}}},\ }\href {http://www.R-project.org} {\emph {\bibinfo {title} {R: A
  Language and Environment for Statistical Computing}}},\ \bibinfo
  {organization} {R Foundation for Statistical Computing},\ \bibinfo {address}
  {Vienna, Austria} (\bibinfo {year} {2008}),\ \bibinfo {note} {{ISBN}
  3-900051-07-0}\BibitemShut {NoStop}%
\bibitem [{\citenamefont {Kaji}\ and\ \citenamefont
  {Morimoto}(2015)}]{kaji201511}%
  \BibitemOpen
  \bibfield  {author} {\bibinfo {author} {\bibfnamefont {D.}~\bibnamefont
  {Kaji}}\ and\ \bibinfo {author} {\bibfnamefont {K.}~\bibnamefont
  {Morimoto}},\ }\href {\doibase https://doi.org/10.1016/j.nima.2015.04.042}
  {\bibfield  {journal} {\bibinfo  {journal} {Nuclear Instruments and Methods
  in Physics Research Section A: Accelerators, Spectrometers, Detectors and
  Associated Equipment}\ }\textbf {\bibinfo {volume} {792}},\ \bibinfo {pages}
  {11 } (\bibinfo {year} {2015})}\BibitemShut {NoStop}%
\bibitem [{\citenamefont {Huang}\ \emph {et~al.}(2017)\citenamefont {Huang},
  \citenamefont {Audi}, \citenamefont {Wang}, \citenamefont {Kondev},
  \citenamefont {Naimi},\ and\ \citenamefont {Xu}}]{huand2017030002}%
  \BibitemOpen
  \bibfield  {author} {\bibinfo {author} {\bibfnamefont {W.}~\bibnamefont
  {Huang}}, \bibinfo {author} {\bibfnamefont {G.}~\bibnamefont {Audi}},
  \bibinfo {author} {\bibfnamefont {M.}~\bibnamefont {Wang}}, \bibinfo {author}
  {\bibfnamefont {F.}~\bibnamefont {Kondev}}, \bibinfo {author} {\bibfnamefont
  {S.}~\bibnamefont {Naimi}}, \ and\ \bibinfo {author} {\bibfnamefont
  {X.}~\bibnamefont {Xu}},\ }\href@noop {} {\bibfield  {journal} {\bibinfo
  {journal} {Chin.\ Phys\ C}\ }\textbf {\bibinfo {volume} {41}},\ \bibinfo
  {pages} {030002} (\bibinfo {year} {2017})}\BibitemShut {NoStop}%
\bibitem [{\citenamefont {Zagrebaev}\ \emph {et~al.}(1999)\citenamefont
  {Zagrebaev}, \citenamefont {Denikin}, \citenamefont {Karpov}, \citenamefont
  {Alekseev}, \citenamefont {Naumenko}, \citenamefont {Rachkov}, \citenamefont
  {Samarin},\ and\ \citenamefont {Saiko}}]{NRVmisc}%
  \BibitemOpen
  \bibfield  {author} {\bibinfo {author} {\bibfnamefont {V.~I.}\ \bibnamefont
  {Zagrebaev}}, \bibinfo {author} {\bibfnamefont {A.~S.}\ \bibnamefont
  {Denikin}}, \bibinfo {author} {\bibfnamefont {A.~V.}\ \bibnamefont {Karpov}},
  \bibinfo {author} {\bibfnamefont {A.~P.}\ \bibnamefont {Alekseev}}, \bibinfo
  {author} {\bibfnamefont {M.~A.}\ \bibnamefont {Naumenko}}, \bibinfo {author}
  {\bibfnamefont {V.~A.}\ \bibnamefont {Rachkov}}, \bibinfo {author}
  {\bibfnamefont {V.~V.}\ \bibnamefont {Samarin}}, \ and\ \bibinfo {author}
  {\bibfnamefont {V.~V.}\ \bibnamefont {Saiko}},\ }\href@noop {} {\enquote
  {\bibinfo {title} {Monte-carlo code of {N}{R}{V}},}\ }\bibinfo {howpublished}
  {\url{http://nrv.jinr.ru/}} (\bibinfo {year} {1999})\BibitemShut {NoStop}%
\bibitem [{\citenamefont {Donets}\ \emph {et~al.}(1966)\citenamefont {Donets},
  \citenamefont {Shchegolev},\ and\ \citenamefont {Ermakov}}]{donets1966}%
  \BibitemOpen
  \bibfield  {author} {\bibinfo {author} {\bibfnamefont {E.~D.}\ \bibnamefont
  {Donets}}, \bibinfo {author} {\bibfnamefont {V.~A.}\ \bibnamefont
  {Shchegolev}}, \ and\ \bibinfo {author} {\bibfnamefont {V.~A.}\ \bibnamefont
  {Ermakov}},\ }\href@noop {} {\bibfield  {journal} {\bibinfo  {journal}
  {Soviet Journal of Nuclear Physics}\ }\textbf {\bibinfo {volume} {2}},\
  \bibinfo {pages} {723} (\bibinfo {year} {1966})}\BibitemShut {NoStop}%
\bibitem [{\citenamefont {Chatillon}\ \emph {et~al.}(2007)\citenamefont
  {Chatillon}, \citenamefont {Theisen}, \citenamefont {Bouchez}, \citenamefont
  {Butler}, \citenamefont {Cl\'ement}, \citenamefont {Dorvaux}, \citenamefont
  {Eeckhaudt}, \citenamefont {Gall}, \citenamefont {G\"orgen}, \citenamefont
  {Grahn}, \citenamefont {Greenlees}, \citenamefont {Herzberg}, \citenamefont
  {He\ss{}berger}, \citenamefont {H\"urstel}, \citenamefont {Jones},
  \citenamefont {Jones}, \citenamefont {Julin}, \citenamefont {Juutinen},
  \citenamefont {Kettunen}, \citenamefont {Khalfallah}, \citenamefont {Korten},
  \citenamefont {Le~Coz}, \citenamefont {Leino}, \citenamefont {Lepp\"anen},
  \citenamefont {Nieminen}, \citenamefont {Pakarinen}, \citenamefont
  {Perkowski}, \citenamefont {Rahkila}, \citenamefont {Rousseau}, \citenamefont
  {Scholey}, \citenamefont {Uusitalo}, \citenamefont {Wilson}, \citenamefont
  {Bonche},\ and\ \citenamefont {Heenen}}]{chatillon2007132503}%
  \BibitemOpen
  \bibfield  {author} {\bibinfo {author} {\bibfnamefont {A.}~\bibnamefont
  {Chatillon}}, \bibinfo {author} {\bibfnamefont {C.}~\bibnamefont {Theisen}},
  \bibinfo {author} {\bibfnamefont {E.}~\bibnamefont {Bouchez}}, \bibinfo
  {author} {\bibfnamefont {P.~A.}\ \bibnamefont {Butler}}, \bibinfo {author}
  {\bibfnamefont {E.}~\bibnamefont {Cl\'ement}}, \bibinfo {author}
  {\bibfnamefont {O.}~\bibnamefont {Dorvaux}}, \bibinfo {author} {\bibfnamefont
  {S.}~\bibnamefont {Eeckhaudt}}, \bibinfo {author} {\bibfnamefont {B.~J.~P.}\
  \bibnamefont {Gall}}, \bibinfo {author} {\bibfnamefont {A.}~\bibnamefont
  {G\"orgen}}, \bibinfo {author} {\bibfnamefont {T.}~\bibnamefont {Grahn}},
  \bibinfo {author} {\bibfnamefont {P.~T.}\ \bibnamefont {Greenlees}}, \bibinfo
  {author} {\bibfnamefont {R.-D.}\ \bibnamefont {Herzberg}}, \bibinfo {author}
  {\bibfnamefont {F.}~\bibnamefont {He\ss{}berger}}, \bibinfo {author}
  {\bibfnamefont {A.}~\bibnamefont {H\"urstel}}, \bibinfo {author}
  {\bibfnamefont {G.~D.}\ \bibnamefont {Jones}}, \bibinfo {author}
  {\bibfnamefont {P.}~\bibnamefont {Jones}}, \bibinfo {author} {\bibfnamefont
  {R.}~\bibnamefont {Julin}}, \bibinfo {author} {\bibfnamefont
  {S.}~\bibnamefont {Juutinen}}, \bibinfo {author} {\bibfnamefont
  {H.}~\bibnamefont {Kettunen}}, \bibinfo {author} {\bibfnamefont
  {F.}~\bibnamefont {Khalfallah}}, \bibinfo {author} {\bibfnamefont
  {W.}~\bibnamefont {Korten}}, \bibinfo {author} {\bibfnamefont
  {Y.}~\bibnamefont {Le~Coz}}, \bibinfo {author} {\bibfnamefont
  {M.}~\bibnamefont {Leino}}, \bibinfo {author} {\bibfnamefont {A.-P.}\
  \bibnamefont {Lepp\"anen}}, \bibinfo {author} {\bibfnamefont
  {P.}~\bibnamefont {Nieminen}}, \bibinfo {author} {\bibfnamefont
  {J.}~\bibnamefont {Pakarinen}}, \bibinfo {author} {\bibfnamefont
  {J.}~\bibnamefont {Perkowski}}, \bibinfo {author} {\bibfnamefont
  {P.}~\bibnamefont {Rahkila}}, \bibinfo {author} {\bibfnamefont
  {M.}~\bibnamefont {Rousseau}}, \bibinfo {author} {\bibfnamefont
  {C.}~\bibnamefont {Scholey}}, \bibinfo {author} {\bibfnamefont
  {J.}~\bibnamefont {Uusitalo}}, \bibinfo {author} {\bibfnamefont {J.~N.}\
  \bibnamefont {Wilson}}, \bibinfo {author} {\bibfnamefont {P.}~\bibnamefont
  {Bonche}}, \ and\ \bibinfo {author} {\bibfnamefont {P.-H.}\ \bibnamefont
  {Heenen}},\ }\href {\doibase 10.1103/PhysRevLett.98.132503} {\bibfield
  {journal} {\bibinfo  {journal} {Phys. Rev. Lett.}\ }\textbf {\bibinfo
  {volume} {98}},\ \bibinfo {pages} {132503} (\bibinfo {year}
  {2007})}\BibitemShut {NoStop}%
\bibitem [{\citenamefont {Oganessian}\ \emph {et~al.}(2001)\citenamefont
  {Oganessian}, \citenamefont {Utyonkov}, \citenamefont {Lobanov},
  \citenamefont {Abdullin}, \citenamefont {Polyakov}, \citenamefont
  {Shirokovsky}, \citenamefont {Tsyganov}, \citenamefont {Mezentsev},
  \citenamefont {Iliev}, \citenamefont {Subbotin}, \citenamefont {Sukhov},
  \citenamefont {Subotic}, \citenamefont {Ivanov}, \citenamefont {Voinov},
  \citenamefont {Zagrebaev}, \citenamefont {Moody}, \citenamefont {Wild},
  \citenamefont {Stoyer}, \citenamefont {Stoyer},\ and\ \citenamefont
  {Lougheed}}]{oganessian2001}%
  \BibitemOpen
  \bibfield  {author} {\bibinfo {author} {\bibfnamefont {Y.~T.}\ \bibnamefont
  {Oganessian}}, \bibinfo {author} {\bibfnamefont {V.~K.}\ \bibnamefont
  {Utyonkov}}, \bibinfo {author} {\bibfnamefont {Y.~V.}\ \bibnamefont
  {Lobanov}}, \bibinfo {author} {\bibfnamefont {F.~S.}\ \bibnamefont
  {Abdullin}}, \bibinfo {author} {\bibfnamefont {A.~N.}\ \bibnamefont
  {Polyakov}}, \bibinfo {author} {\bibfnamefont {I.~V.}\ \bibnamefont
  {Shirokovsky}}, \bibinfo {author} {\bibfnamefont {Y.~S.}\ \bibnamefont
  {Tsyganov}}, \bibinfo {author} {\bibfnamefont {A.~N.}\ \bibnamefont
  {Mezentsev}}, \bibinfo {author} {\bibfnamefont {S.}~\bibnamefont {Iliev}},
  \bibinfo {author} {\bibfnamefont {V.~G.}\ \bibnamefont {Subbotin}}, \bibinfo
  {author} {\bibfnamefont {A.~M.}\ \bibnamefont {Sukhov}}, \bibinfo {author}
  {\bibfnamefont {K.}~\bibnamefont {Subotic}}, \bibinfo {author} {\bibfnamefont
  {O.~V.}\ \bibnamefont {Ivanov}}, \bibinfo {author} {\bibfnamefont {A.~N.}\
  \bibnamefont {Voinov}}, \bibinfo {author} {\bibfnamefont {V.~I.}\
  \bibnamefont {Zagrebaev}}, \bibinfo {author} {\bibfnamefont {K.~J.}\
  \bibnamefont {Moody}}, \bibinfo {author} {\bibfnamefont {J.~F.}\ \bibnamefont
  {Wild}}, \bibinfo {author} {\bibfnamefont {N.~J.}\ \bibnamefont {Stoyer}},
  \bibinfo {author} {\bibfnamefont {M.~A.}\ \bibnamefont {Stoyer}}, \ and\
  \bibinfo {author} {\bibfnamefont {R.~W.}\ \bibnamefont {Lougheed}},\ }\href
  {\doibase 10.1103/PhysRevC.64.054606} {\bibfield  {journal} {\bibinfo
  {journal} {Phys. Rev. C}\ }\textbf {\bibinfo {volume} {64}},\ \bibinfo
  {pages} {054606} (\bibinfo {year} {2001})}\BibitemShut {NoStop}%
\bibitem [{\citenamefont {Wang}\ \emph {et~al.}(2017)\citenamefont {Wang},
  \citenamefont {Audi}, \citenamefont {Kondev}, \citenamefont {Huang},
  \citenamefont {Naimi},\ and\ \citenamefont {Xu}}]{ame2016:2}%
  \BibitemOpen
  \bibfield  {author} {\bibinfo {author} {\bibfnamefont {M.}~\bibnamefont
  {Wang}}, \bibinfo {author} {\bibfnamefont {G.}~\bibnamefont {Audi}}, \bibinfo
  {author} {\bibfnamefont {F.~G.}\ \bibnamefont {Kondev}}, \bibinfo {author}
  {\bibfnamefont {W.~J.}\ \bibnamefont {Huang}}, \bibinfo {author}
  {\bibfnamefont {S.}~\bibnamefont {Naimi}}, \ and\ \bibinfo {author}
  {\bibfnamefont {X.}~\bibnamefont {Xu}},\ }\href@noop {} {\bibfield  {journal}
  {\bibinfo  {journal} {Chinese physics C}\ }\textbf {\bibinfo {volume} {41}},\
  \bibinfo {pages} {30003} (\bibinfo {year} {2017})}\BibitemShut {NoStop}%
\bibitem [{\citenamefont {He{\ss}berger}\ \emph {et~al.}(2010)\citenamefont
  {He{\ss}berger}, \citenamefont {Antalic}, \citenamefont {Sulignano},
  \citenamefont {Ackermann}, \citenamefont {Heinz}, \citenamefont {Hofmann},
  \citenamefont {Kindler}, \citenamefont {Khuyagbaatar}, \citenamefont
  {Kojouharov}, \citenamefont {Kuusiniemi}, \citenamefont {Leino},
  \citenamefont {Lommel}, \citenamefont {Mann}, \citenamefont {Nishio},
  \citenamefont {Popeko}, \citenamefont {Saro}, \citenamefont {Streicher},
  \citenamefont {Uusitalo}, \citenamefont {Venhart},\ and\ \citenamefont
  {Yeremin}}]{hessberger201055}%
  \BibitemOpen
  \bibfield  {author} {\bibinfo {author} {\bibfnamefont {F.~P.}\ \bibnamefont
  {He{\ss}berger}}, \bibinfo {author} {\bibfnamefont {S.}~\bibnamefont
  {Antalic}}, \bibinfo {author} {\bibfnamefont {B.}~\bibnamefont {Sulignano}},
  \bibinfo {author} {\bibfnamefont {D.}~\bibnamefont {Ackermann}}, \bibinfo
  {author} {\bibfnamefont {S.}~\bibnamefont {Heinz}}, \bibinfo {author}
  {\bibfnamefont {S.}~\bibnamefont {Hofmann}}, \bibinfo {author} {\bibfnamefont
  {B.}~\bibnamefont {Kindler}}, \bibinfo {author} {\bibfnamefont
  {J.}~\bibnamefont {Khuyagbaatar}}, \bibinfo {author} {\bibfnamefont
  {I.}~\bibnamefont {Kojouharov}}, \bibinfo {author} {\bibfnamefont
  {P.}~\bibnamefont {Kuusiniemi}}, \bibinfo {author} {\bibfnamefont
  {M.}~\bibnamefont {Leino}}, \bibinfo {author} {\bibfnamefont
  {B.}~\bibnamefont {Lommel}}, \bibinfo {author} {\bibfnamefont
  {R.}~\bibnamefont {Mann}}, \bibinfo {author} {\bibfnamefont {K.}~\bibnamefont
  {Nishio}}, \bibinfo {author} {\bibfnamefont {A.~G.}\ \bibnamefont {Popeko}},
  \bibinfo {author} {\bibfnamefont {S.}~\bibnamefont {Saro}}, \bibinfo {author}
  {\bibfnamefont {B.}~\bibnamefont {Streicher}}, \bibinfo {author}
  {\bibfnamefont {J.}~\bibnamefont {Uusitalo}}, \bibinfo {author}
  {\bibfnamefont {M.}~\bibnamefont {Venhart}}, \ and\ \bibinfo {author}
  {\bibfnamefont {A.~V.}\ \bibnamefont {Yeremin}},\ }\href@noop {} {\bibfield
  {journal} {\bibinfo  {journal} {Eur.\ Phys.\ J.\ A}\ }\textbf {\bibinfo
  {volume} {43}},\ \bibinfo {pages} {55} (\bibinfo {year} {2010})}\BibitemShut
  {NoStop}%
\bibitem [{\citenamefont {Rutz}\ \emph {et~al.}(1997)\citenamefont {Rutz},
  \citenamefont {Bender}, \citenamefont {B\"urvenich}, \citenamefont
  {Schilling}, \citenamefont {Reinhard}, \citenamefont {Maruhn},\ and\
  \citenamefont {Greiner}}]{rutz1997238}%
  \BibitemOpen
  \bibfield  {author} {\bibinfo {author} {\bibfnamefont {K.}~\bibnamefont
  {Rutz}}, \bibinfo {author} {\bibfnamefont {M.}~\bibnamefont {Bender}},
  \bibinfo {author} {\bibfnamefont {T.}~\bibnamefont {B\"urvenich}}, \bibinfo
  {author} {\bibfnamefont {T.}~\bibnamefont {Schilling}}, \bibinfo {author}
  {\bibfnamefont {P.-G.}\ \bibnamefont {Reinhard}}, \bibinfo {author}
  {\bibfnamefont {J.~A.}\ \bibnamefont {Maruhn}}, \ and\ \bibinfo {author}
  {\bibfnamefont {W.}~\bibnamefont {Greiner}},\ }\href {\doibase
  10.1103/PhysRevC.56.238} {\bibfield  {journal} {\bibinfo  {journal} {Phys.
  Rev. C}\ }\textbf {\bibinfo {volume} {56}},\ \bibinfo {pages} {238} (\bibinfo
  {year} {1997})}\BibitemShut {NoStop}%
\bibitem [{\citenamefont {Duflo}\ and\ \citenamefont
  {Zuker}(1995)}]{duflo1995R23}%
  \BibitemOpen
  \bibfield  {author} {\bibinfo {author} {\bibfnamefont {J.}~\bibnamefont
  {Duflo}}\ and\ \bibinfo {author} {\bibfnamefont {A.}~\bibnamefont {Zuker}},\
  }\href {\doibase 10.1103/PhysRevC.52.R23} {\bibfield  {journal} {\bibinfo
  {journal} {Phys. Rev. C}\ }\textbf {\bibinfo {volume} {52}},\ \bibinfo
  {pages} {R23} (\bibinfo {year} {1995})}\BibitemShut {NoStop}%
\bibitem [{\citenamefont {M{\"o}ller}\ \emph {et~al.}(2016)\citenamefont
  {M{\"o}ller}, \citenamefont {Sierk}, \citenamefont {Ichikawa},\ and\
  \citenamefont {Sagawa}}]{moller20161}%
  \BibitemOpen
  \bibfield  {author} {\bibinfo {author} {\bibfnamefont {P.}~\bibnamefont
  {M{\"o}ller}}, \bibinfo {author} {\bibfnamefont {A.}~\bibnamefont {Sierk}},
  \bibinfo {author} {\bibfnamefont {T.}~\bibnamefont {Ichikawa}}, \ and\
  \bibinfo {author} {\bibfnamefont {H.}~\bibnamefont {Sagawa}},\ }\href@noop {}
  {\bibfield  {journal} {\bibinfo  {journal} {Atomic Data and Nuclear Data
  Tables}\ }\textbf {\bibinfo {volume} {109}},\ \bibinfo {pages} {1} (\bibinfo
  {year} {2016})}\BibitemShut {NoStop}%
\bibitem [{\citenamefont {Wang}\ and\ \citenamefont
  {Liu}(2011)}]{wang2011051303}%
  \BibitemOpen
  \bibfield  {author} {\bibinfo {author} {\bibfnamefont {N.}~\bibnamefont
  {Wang}}\ and\ \bibinfo {author} {\bibfnamefont {M.}~\bibnamefont {Liu}},\
  }\href@noop {} {\bibfield  {journal} {\bibinfo  {journal} {Phys. Rev. C}\
  }\textbf {\bibinfo {volume} {84}},\ \bibinfo {pages} {051303} (\bibinfo
  {year} {2011})}\BibitemShut {NoStop}%
\bibitem [{\citenamefont {Goriely}\ \emph {et~al.}(2016)\citenamefont
  {Goriely}, \citenamefont {Chamel},\ and\ \citenamefont
  {Pearson}}]{goriely2016034337}%
  \BibitemOpen
  \bibfield  {author} {\bibinfo {author} {\bibfnamefont {S.}~\bibnamefont
  {Goriely}}, \bibinfo {author} {\bibfnamefont {N.}~\bibnamefont {Chamel}}, \
  and\ \bibinfo {author} {\bibfnamefont {J.~M.}\ \bibnamefont {Pearson}},\
  }\href {\doibase 10.1103/PhysRevC.93.034337} {\bibfield  {journal} {\bibinfo
  {journal} {Phys. Rev. C}\ }\textbf {\bibinfo {volume} {93}},\ \bibinfo
  {pages} {034337} (\bibinfo {year} {2016})}\BibitemShut {NoStop}%
\bibitem [{\citenamefont {Koura}\ \emph {et~al.}(2005)\citenamefont {Koura},
  \citenamefont {Tachibana}, \citenamefont {Uno},\ and\ \citenamefont
  {Yamada}}]{koura2005305}%
  \BibitemOpen
  \bibfield  {author} {\bibinfo {author} {\bibfnamefont {H.}~\bibnamefont
  {Koura}}, \bibinfo {author} {\bibfnamefont {T.}~\bibnamefont {Tachibana}},
  \bibinfo {author} {\bibfnamefont {M.}~\bibnamefont {Uno}}, \ and\ \bibinfo
  {author} {\bibfnamefont {M.}~\bibnamefont {Yamada}},\ }\href@noop {}
  {\bibfield  {journal} {\bibinfo  {journal} {Progress of Theoretical Physics}\
  }\textbf {\bibinfo {volume} {113}},\ \bibinfo {pages} {305} (\bibinfo {year}
  {2005})}\BibitemShut {NoStop}%
\bibitem [{\citenamefont {Wang}\ \emph {et~al.}(2014)\citenamefont {Wang},
  \citenamefont {Liu}, \citenamefont {Wu},\ and\ \citenamefont
  {Meng}}]{wang2014215}%
  \BibitemOpen
  \bibfield  {author} {\bibinfo {author} {\bibfnamefont {N.}~\bibnamefont
  {Wang}}, \bibinfo {author} {\bibfnamefont {M.}~\bibnamefont {Liu}}, \bibinfo
  {author} {\bibfnamefont {X.}~\bibnamefont {Wu}}, \ and\ \bibinfo {author}
  {\bibfnamefont {J.}~\bibnamefont {Meng}},\ }\href@noop {} {\bibfield
  {journal} {\bibinfo  {journal} {Physics Letters B}\ }\textbf {\bibinfo
  {volume} {734}},\ \bibinfo {pages} {215} (\bibinfo {year}
  {2014})}\BibitemShut {NoStop}%
\end{thebibliography}%

\end{document}